\documentclass[paper]{JFM-FLM_Au}

\graphicspath{ {figures/} }

\usepackage{graphicx}
\usepackage{newtxtext}
\usepackage{newtxmath}
\usepackage{natbib}
\usepackage[dvipsnames]{xcolor}
\usepackage{hyperref}
\hypersetup{  colorlinks,
              linktoc         = page, 
              urlcolor        = {green!80!blue},
              linkcolor       = {orange!60!black}, 
              urlcolor        = {magenta!80!black}, 
              citecolor       = {magenta!80!black}, 
              anchorcolor     = {yellow}
}

\usepackage{bm}
\newcommand{\tp}{{\mathrm{tp}}}
\newcommand{\dpp}{{\mathrm{dpp}}}
\newcommand{\dpr}{{\mathrm{dpr}}}
\newcommand{\far}{{\mathrm{far}}}

\newcommand{\RomanNumeralCaps}[1]

\def\grad{\bnabla}

\lefttitle{Sam~Patrick et al.}
\righttitle{When is a sloshing vortex an analogue black hole bomb?}

\title{When is a sloshing vortex an analogue black hole bomb?}

\author{Sam~Patrick\aff{1}, Leonardo Solidoro\aff{2,3}, Maur\'icio Richartz\aff{4}, Pietro Smaniotto\aff{2,3}, Patrik \v{S}van\v{c}ara\aff{2,3}\corresp{Present address: Institut N\'{e}el, CNRS, 25 avenue des Martyrs, 38042 Grenoble, France}, Silke Weinfurtner\aff{2,3,5} \and Ruth Gregory\aff{1,6}
}

\affiliation{\aff{1}Department of Physics, King’s College London, University of London, Strand, London, WC2R 2LS, UK
\aff{2}School of Mathematical Sciences, University of Nottingham, University Park, Nottingham, NG7 2RD, UK
\aff{3}Nottingham Centre of Gravity, University of Nottingham,
University Park, Nottingham NG7 2RD, UK
\aff{4}Centro de Matem\'atica, Computa\c c\~ao e Cogni\c c\~ao,
Universidade Federal do ABC (UFABC), 09210-170 Santo Andr\'e, S\~ao Paulo, Brazil
\aff{5}Centre for the Mathematics and Theoretical Physics of Quantum Non-Equilibrium Systems, University of Nottingham, Nottingham, NG7 2RD, UK
\aff{6}Perimeter Institute, 31 Caroline Street North, Waterloo, ON, N2L 2Y5, Canada
}

\corresau{Sam~Patrick, samuel\_christian.patrick@kcl.ac.uk}

\begin{document}
\maketitle

\begin{abstract}
Draining vortices provide a powerful platform for simulating black hole phenomena in tabletop experiments. In realistic fluid systems confined within a finite container, low-frequency waves amplified by the vortex are reflected at the walls, rendering the system unstable. This process, known in the gravitational context as the black hole bomb, manifests as a sloshing motion of the free surface. The analogy, however, becomes more nuanced when a realistic vortex core with a non-singular vorticity distribution is considered. We investigate this by analysing a non-draining Rankine vortex in the shallow-water and inviscid limits. At low circulation, the sloshing corresponds to an instability of the vorticity field, whereas at high circulation where fluid is expelled from the vortex core, the destabilising mechanism coincides with that of the black hole bomb. Our variational framework distinguishes the energetic contributions of vorticity and irrotational perturbations, offering new insight into the rotating-polygons instability reported by, e.g. \citet{jansson2006polygons}. From the analogue-gravity perspective, we identify hollow core vortices as an optimal regime for exploring black-hole-like instabilities in fluids.
\end{abstract}



\section{Introduction}
\label{sec:intro}

Rotating free-surface flows provide a rich arena for studying nonlinear hydrodynamic instabilities and wave-vortex interactions. A canonical example is the swirling flow in a cylindrical container driven by rotation of the bottom plate, which exhibits spontaneous symmetry breaking of the free surface into polygonal, sloshing, and switching states \citep{vatistas1990note,jansson2006polygons,suzuki2006surface,tasaka2009flow,bergmann2011polygon,iga2014various}. The fluid-dynamical mechanisms underlying these phenomena have been progressively clarified through linear and weakly nonlinear analyses. The onset of rotating polygonal patterns was first interpreted by \citet{tophoj2013rotating} as a resonance between centrifugal and gravity surface waves in a potential vortex flow model, a view later elaborated by \citet{mougel2017instabilities} using global stability and Wentzel-Kramers-Brillouin (WKB) methods. Subsequent refinements introduced Rankine-type vortices incorporating a central vorticity-containing core to explain lower-rotation (``wet'') regimes where the fluid entirely covers the bottom part of the container~\citep{fabre2014generation,mougel2014waves}.
These studies identified additional resonance mechanisms involving Kelvin-Kirchhoff or Rossby-like waves, accounting for the sloshing and switching motions observed experimentally. Together, this body of work established that a broad class of free-surface instabilities in rotating vortices can be understood as interactions among distinct surface-wave families governed by the base-flow structure and Froude number. In the present study, we revisit these dynamics from a complementary theoretical perspective, employing a variational formulation motivated by analogue gravity methods to describe the coupling between free-surface and vortex-core oscillations.

This approach is based on a close analogy between inhomogeneous flows and wave dynamics in curved spacetime geometries. The relationship was discovered by \citet{unruh1981experimental}, who showed that sound waves in an irrotational flow obey an equation identical to that of a scalar field in a curved spacetime, enabling the study of the then-controversial phenomenon of black hole evaporation \citep{hawking1974black}. A key feature of this framework (known as analogue gravity) is that the fluid is assumed irrotational at the outset. The velocity field $\mathbf{v}=\grad\phi$ is then determined by a single scalar degree of freedom $\phi$, called the velocity potential, whose perturbations are identified with the scalar field in the gravitational analogy. Since this seminal work, the analogy has been extended to a wide range of systems \citep[see, e.g.][]{barcelo2011analogue,jacquet2020next}, including long-wavelength surface-gravity waves~\citep{schutzhold2002gravity}, leading to the measurement of experimental signatures associated with Hawking radiation in both classical~\citep{weinfurtner2011measurement,euve2016observation} and quantum fluids~\citep{steinhauer2016observation,munoz2019observation}.

Different effective curved spacetimes can be realised by suitably engineering the fluid flow. A steady draining vortex flow \citep{andersen2003anatomy}, for instance, emulates key kinematic features of a rotating black hole spacetime~\citep[see, e.g.][]{visser1998acoustic,berti2004quasinormal,dolan2012resonances}. In such flows, long-wavelength surface waves encounter a region where the radial velocity of the fluid exceeds their intrinsic propagation speed, thereby forming the analogue of a black hole horizon. An ergoregion (a region where no disturbance can remain stationary with respect to the laboratory frame) also appears where the combined radial and azimuthal flow speeds surpass the wave speed. The presence of an ergoregion gives rise to an amplification process known as rotational superradiance~\citep{basak2003superresonance,richartz2015rotating,Brito:2015oca}. The mechanism relies on the fact that perturbations of the velocity potential can possess negative energy inside this region (with respect to the laboratory frame). Consequently, when an incoming wave scatters off the vortex, the absorption of its negative-energy component by the analogue horizon leads to an amplified reflection. This energy-extraction mechanism, fundamental to gravitational physics, allows a rotating black hole to shed angular momentum~\citep{page1976particle}. To connect back to fluid dynamics, superradiance, more commonly known as over-reflection in this context, is the mechanism at the heart of the rotating polygons instability of swirling free surface flows \citep{mougel2017instabilities}, as we now discuss.

\begin{figure}
\centering
\includegraphics[scale=1]{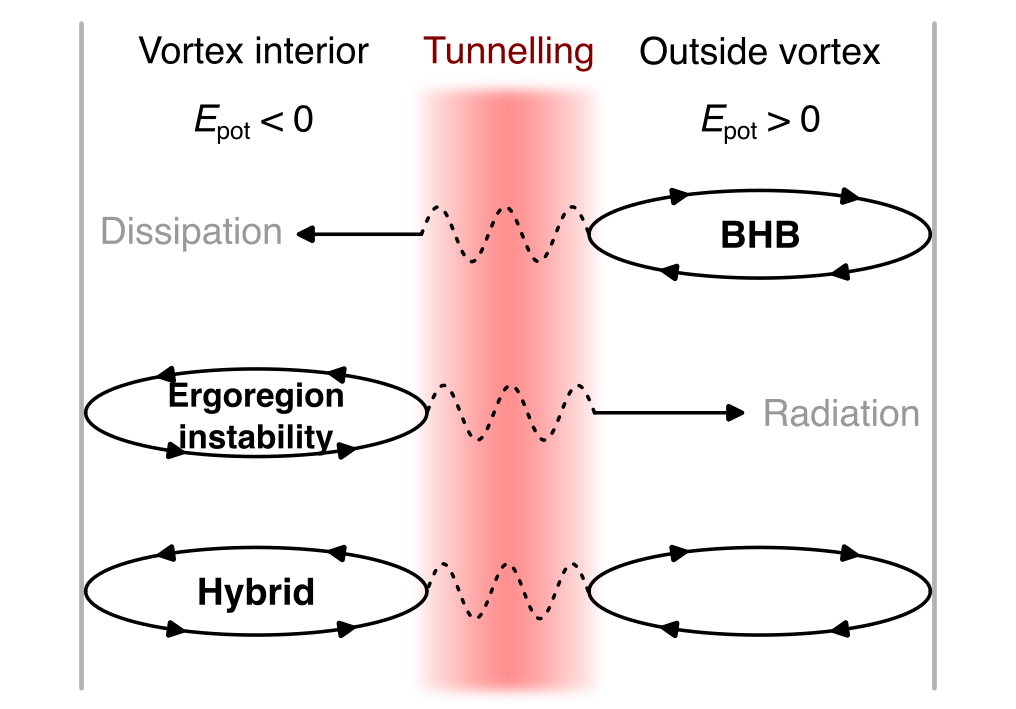}
\caption{Various types of superradiant instabilities in irrotational swirling flows. Top: the black hole bomb (BHB) occurs as a positive-energy state which is trapped outside the vortex, between the effective potential barrier representing the analogue ergoregion (red; see \S~\ref{ssec:lagr-perturbations}) and the outer boundary. It grows because negative energy is transmitted into the interior and dissipated, e.g. absorbed at the event horizon. Middle: the ergoregion instability is a negative-energy state in the interior region, which grows when positive energy is radiated to infinity. Bottom: if the system is closed on both ends, a hybrid instability occurs when the inner and outer states resonate with each other. Although the three types of instability differ by which region of space they are localised in, they all rely on energy transfer across the ergoregion.}
\label{fig:1}
\end{figure}

When superradiant scattering is coupled with a mechanism that confines the waves, repeated amplification can occur, giving rise to instability. The specific form of this superradiant instability depends on where the trapping occurs. For example, if a black hole is surrounded by a reflective boundary, the amplified waves can be repeatedly scattered between the boundary and the ergoregion, producing an exponentially growing perturbation outside the black hole~\citep{Cardoso:2004nk}. This phenomenon, termed the black hole bomb (BHB) by \citet{press1972floating}, represents the canonical example of a superradiant instability and is illustrated in the top part of Fig.~\ref{fig:1}. The instability comprises a growing standing wave with positive energy, and a negative-energy propagating wave in the ergoregion which is dissipated by an absorption mechanism. In the case of a black hole, this is provided by the event horizon, representing a perfect absorber, as per its definition. An instability of the BHB type can still occur if this assumption is relaxed and the inner boundary is only partially reflective~\citep{Cardoso:2016zvz,torres2022imperfect,patrick2024quantum}.
The conditions for a BHB instability are naturally realised around a draining vortex confined within a cylindrical container, as studied in water by \citet{andersen2003anatomy} and more recently in superfluid helium by \citet{yano2018observation,matsumura2019observation,obara2021vortex,svancara2024rotating,smaniotto2025}. We note that the inherently low-dissipation of superfluids makes them natural candidates to investigate superradiant instabilities in the deep nonlinear regime, since viscous damping in classical fluids tends to suppress their growth rates~\citep{patrick2024primer}.

A second type of instability can occur even if the system is open, i.e. there is no reflective outer boundary, provided there is no (or minimal) dissipation inside the ergoregion. This scenario (which is in a sense the mirror image of the BHB) involves a negative-energy standing wave in the ergoregion radiating positive energy to infinity, and is illustrated in the middle part Fig.~\ref{fig:1}. This is known as the ergoregion instability~\citep{Friedman:1978ygc,comins_schutz} and it arises around compact gravitational objects which lack a horizon, e.g. neutron stars~\citep{kokkotas2004w}, hypothetical boson stars~\citep{cardoso2008ergoregion}, and in classical fluids for non-draining vortices ~\citep{oliveira2014ergoregion,oliveira2018ergoregion}. Furthermore, in quantum fluids the ergoregion instability drives the decay of multiply-quantised vortices into clusters of single vortex quanta~\citep{isoshima2007spontaneous,okano2007splitting,takeuchi2018doubly,patrick2022quantum}.

Finally, there is a hybrid instability which occurs when there is a reflective outer boundary and no dissipative mechanism in the ergoregion. This can be viewed as a combination of the BHB and the ergoregion instability in the sense that there are two standing waves (inside and outside the ergoregion respectively) which must be frequency matched to each other, indicated at the bottom part of Fig.~\ref{fig:1}. Due to the frequency matching, this instability only occurs in certain regions of the parameter space, placing constraints on, for example, the splitting patterns of quantum vortices in confined systems \citep{giacomelli2020ergoregion,patrick2022origin}. More relevant to our study here, the analysis of \citet{mougel2017instabilities} reveals that this hybrid instability is responsible for the rotating polygons instability mentioned above.

Note that the instability landscape we have just outlined applies only to an irrotational fluid, where the analogy with the corresponding gravitational processes is clear. In fluids with vorticity, however, the dynamics of free-surface swirling flows is richer due to the presence of additional, vorticity-carrying degrees of freedom that arise inside the vortex core \citep{fabre2014generation,mougel2014waves}. Since these excitations can also carry negative energy, they can couple to standing waves outside the vortex core to produce an instability similar to the hybrid one shown in Fig.~\ref{fig:1}. However, as we soon show, this instability can exist even when there is no ergoregion, making it somewhat different from the superradiant instability in gravitational physics. Our aim in this article is to clarify the physical origin of this vorticity instability and to understand its transition into the superradiant instability as we vary the speed of the background flow.

\section{Theoretical modelling}

Since the irrotational assumption plays a key role in identifying an effective spacetime metric, treatments of vorticity within the analogue gravity framework are usually approximate in nature.
For example, \citet{patrick2018black,oliveira2024ergoregion} studied irrotational perturbations in backgrounds with vorticity, whereas \citep{churilov2019scattering} allowed for vorticity perturbations but treated the background as irrotational.
We also mention that vorticity has a been studied in 1D analogues with boundary shear flows \citep{biondi2024vortical} and Bose-Einstein condensates coupled to the electromagnetic field \citep{cropp2016vorticity,liberati2019vorticity}.
A full framework that treats both perturbations and background with vorticity (whilst maintaining the connection to analogue gravity) was put forward by \citet{bergliaffa2004wave}, and this framework was applied to study amplification of sound waves around a non-draining vortex by \citet{richartz2009generalized}.
In the present work, we extend this framework to study shallow surface waves in an inviscid fluid.

\subsection{Variational framework}
\label{sec:lagrangian}
To describe an incompressible and inviscid fluid flow with vorticity, we use the Clebsch decomposition to write the three-dimensional (3D) velocity field of the fluid as
\begin{equation} \label{vel_general}
    \mathbf{v} = \grad\phi + \alpha\grad\beta,
\end{equation}
where the three scalar functions $\alpha$, $\beta$, and $\phi$ are called Clebsch potentials.
When the flow is potential, the velocity field satisfies $\grad\times\mathbf{v}=0$, implying $\alpha=\beta=0$. In this case, the velocity field contains only a single degree of freedom, the velocity potential $\phi$.
On the other hand, for flows with vorticity, $\alpha$ and $\beta$ are non-zero and $\mathbf{v}$ contains three degrees of freedom.
The associated vorticity vector is given by
\begin{equation} \label{vort}
    \bm{\zeta} = \grad\times\mathbf{v} = \grad\alpha\times\grad\beta.
\end{equation}
Employing a Cartesian coordinate system $(t,x,y,z)$ and assuming a gravitational acceleration $\mathbf{g}  =  -g \, \hat{\mathbf{e}}_z$, it can be shown that the equations of motion for $\phi$, $\alpha$, $\beta$, and the pressure $P$, are~\citep{bergliaffa2004wave}
\begin{subequations}
\begin{align}
    \grad\cdot\mathbf{v} &= 0, \label{divfree} \\
    \partial_t\alpha + \grad\cdot(\mathbf{v}\alpha) &= 0, \label{alpha_eom} \\
    \partial_t\beta + \mathbf{v}\cdot\grad\beta &= 0, \label{beta_eom} \\
    \partial_t {\phi} + \alpha \partial_t {\beta} + \frac{1}{2}\mathbf{v}^2 + \frac{P}{\rho} + gz &= 0, \label{bern}
\end{align}
\end{subequations}
where $\rho$ is the density of the fluid and $\mathbf{v}$ a function of $\phi,\alpha,\beta$ given in \eqref{vel_general}.
Taking the time derivative of $\mathbf{v}$ and substituting the four equations above, we obtain the Euler equation,
\begin{equation} \label{euler}
    (\partial_t+\mathbf{v}\cdot\grad)\mathbf{v} + \frac{\grad P}{\rho} - \mathbf{g} = 0.
\end{equation}
We are interested in the dynamics of a free surface located at $z=h(x,y,t)$. Neglecting surface tension and ambient pressure, the free surface boundary condition, i.e.~$P=0$ at $z=h$, converts \eqref{bern} into an equation for the unknown variable $h$,
\begin{equation} \label{bern_surf}
    \left(\partial_t{\phi} + \alpha \partial_t {\beta} + \frac{1}{2}\mathbf{v}^2\right)_{z=h} + gh = 0.
\end{equation}
The system admits a variational formulation by identifying the pressure in \eqref{bern} as the Lagrangian density~\citep{luke1967variational}. The corresponding action is,
\begin{equation} \label{action1}
    S = -\int dt\,d^2\mathbf{x}_\parallel\int^h_0dz\,\rho\left(\partial_t{\phi} + \alpha\partial_t{\beta} + \frac{1}{2}\mathbf{v}^2 + gz\right),
\end{equation}
where $\mathbf{x}_\parallel = (x,y)$. Minimising $S$ is equivalent to local thermodynamic equilibrium since variations in the pressure and thermodynamic energy are related by $dE = -PdV$. We note that the reason we have been able to write down an action principle is because viscosity has been neglected, which renders the dynamics conservative. This has two main consequences which we soon exploit in our analysis: firstly, the background vortex has a simple steady solution and, secondly, there is a notion of energy conservation for the perturbations. Neglecting viscosity in this way amounts to the assumption that the timescale associated with viscous diffusion is much longer than other timescales in the problem.

Stationarity of the action under independent variations of $\phi$, $\alpha$, $\beta$, and $h$ yields the equations of motion \eqref{divfree}, \eqref{alpha_eom}, \eqref{beta_eom} and \eqref{bern_surf}, respectively.
The boundary conditions are obtained by requiring $\delta S=0$ when the boundary fields are varied independently,
\begin{subequations}
\begin{align}
    (\delta\phi+\alpha\delta\beta)_{z=h}: && v_z(z=h) &= \partial_t h + \mathbf{v}_\parallel\cdot\grad_\parallel h, \\
    (\delta\phi+\alpha\delta\beta)_{z=0}: && v_z(z=0) &= 0, \\
    (\delta\phi+\alpha\delta\beta)_{\mathbf{x}_\parallel = \mathbf{x}_\parallel^B}: && \hat{\mathbf{n}}_B\cdot\mathbf{v}_\parallel &= 0,
\end{align}
\end{subequations}
where $\hat{\mathbf{n}}_B$ is the normal vector to the boundary, which we assume to be a hard wall at $\mathbf{x}_\parallel = \mathbf{x}_\parallel^B $, and $\mathbf{v}_\parallel$ denotes the components of $\mathbf{v}$ in the $(x,y)$-plane.

Considerable simplification occurs in the shallow water regime, which considers the fluid depth $h$ to be much smaller than the wavelength of any relevant perturbation.
Assuming that the flow is essentially two-dimensional (2D),
the action \eqref{action1} becomes,
\begin{equation} \label{action2}
    S = -\int d^2\mathbf{x}_\parallel \left(h\partial_t{\phi} + h\alpha\partial_t{\beta} + \frac{1}{2}h\mathbf{v}_\parallel^2 + \frac{1}{2}gh^2\right).
\end{equation}
To simplify our notation, from here on we drop the subscript $\parallel$ since the action above and the quantities that define it are now independent of $z$. Note that we have also dropped the prefactor $\rho$ in the expression above since the density is assumed constant and does not contribute to the dynamics.
Variation of \eqref{action2} leads to the shallow water equations of motion, which are \eqref{alpha_eom}, \eqref{beta_eom} and \eqref{bern_surf} restricted to the $(x,y)$-plane, together with a continuity equation for the height field,
\begin{equation}
    \partial_t h + \grad\cdot(h\mathbf{v}) = 0.
\end{equation}

We are interested in studying the perturbations of a stationary background fluid flow containing a vortex at the coordinate origin.
To this end, we perturb the fluid variables in the following way,
\begin{subequations}
\begin{align}
    h = & \  h_0 + \epsilon h_1 + \mathcal{O}(\epsilon^2), \\
    \phi = & \  \phi_0 + \epsilon \phi_1 + \mathcal{O}(\epsilon^2), \\
    \alpha = & \  \alpha_0 + \epsilon \alpha_1 + \mathcal{O}(\epsilon^2), \\
    \beta = & \  \beta_0 + \epsilon \beta_1 + \mathcal{O}(\epsilon^2), 
\end{align}
\end{subequations}
so that
\begin{subequations}
\begin{align}
    \mathbf{v} = & \ \mathbf{v}_0 + \epsilon  \mathbf{v}_1 + \mathcal{O}(\epsilon^2), \\
    \bm{\zeta} = & \ \bm{\zeta}_0 + \epsilon  \bm{\zeta}_1 + \mathcal{O}(\epsilon^2), 
\end{align}
\end{subequations}
where the infinitesimal parameter $\epsilon$ is an order-counting parameter for the perturbations. We then
expand the action \eqref{action2} in powers of $\epsilon$.
At $\mathcal{O}(\epsilon)$, we obtain the shallow water system of equations for the background quantities.
At $\mathcal{O}(\epsilon^2)$, we obtain an action for the linear perturbations,
\begin{equation} \label{action_pert}
\begin{split}
    S_2 = & \ -\int d^2\mathbf{x} \Big[h_1D_t\phi_1 + \left(\alpha_0 h_1+\alpha_1h_0\right) D_t\beta_1 + \frac{1}{2}gh_1^2+\frac{1}{2}h_0|\grad\phi_1|^2 + h_0\alpha_1\grad\phi_1\cdot\grad\beta_0 +\\
    & \quad + h_0\alpha_0\grad\phi_1\cdot\grad\beta_1 + \frac{1}{2}h_0|\grad\beta_0|^2\alpha_1^2 + \frac{1}{2}h_0\alpha_0^2|\grad\beta_1|^2 + h_0\alpha_0\alpha_1\grad\beta_0\cdot\grad\beta_1 \Big],
\end{split}
\end{equation}
where $D_t = \partial_t+\mathbf{v}_0\cdot\grad$ denotes the material derivative,
which can be used to deduce the equations of motion for the quadruplet of perturbative fields
\begin{equation} \label{quadruplet}
X_1=(\phi_1,h_1,\alpha_1,\beta_1).
\end{equation}

As noted in \citet{bergliaffa2004wave}, the decomposition of $\mathbf{v}$ into $\phi$, $\alpha$ and $\beta$ is not unique, implying a residual gauge freedom. It is useful to define two new fields, $\psi_1$ and $\bm{\xi}_1$, in terms of the Clebsch potentials and its perturbations following the relations below, 
\begin{subequations} \label{gauge_inv}
\begin{align} 
    \psi_1 &= \phi_1 + \alpha_0\beta_1, \\
    \bm{\xi}_1 &= \alpha_1\grad\beta_0-\beta_1\grad\alpha_0,
\end{align}
\end{subequations}
such that the perturbed velocity field is given by
\begin{equation} \label{v1_eq}
    \mathbf{v}_1 = \grad\psi_1 + \bm{\xi}_1.
\end{equation}
The linear equations of motion, obtained from the action \eqref{action_pert}, can then be expressed in the form,
\begin{subequations} \label{lin_eqs}
\begin{align}
    \partial_th_1 + \grad\cdot(h_1\mathbf{v}_0 + h_0\mathbf{v}_1) = & \ 0, \label{h1_eom} \\
    D_t\psi_1 + gh_1 = & \ 0, \label{psi1_eom} \\
    D_t\bm{\xi}_1 + \bm{\xi}_1\cdot\grad\mathbf{v}_0 =& \grad\psi_1\times \bm{\zeta}_0. \label{xi_eom}
\end{align}
\end{subequations}
Note that when varying the action, one should express $\mathbf{v}_1$ using Eq.~\eqref{vel_general} expanded to linear order in $\phi_1,\alpha_1,\beta_1$ and then vary with respect to these quantities (as well as $h_1$). The resulting equations are then conveniently expressed in the form \eqref{lin_eqs} using the definitions in \eqref{gauge_inv} \citep[see][for details]{bergliaffa2004wave}.

Since $\psi_1$ appears in an equation with only physical (measurable) quantities (namely $\mathbf{v}_0$ and $h_1$), $\psi_1$ itself is a gauge invariant quantity.
Similarly, by \eqref{v1_eq} we have $\bm{\xi}_1 = \mathbf{v}_1-\grad\psi_1$.
Therefore, since $\mathbf{v}_1$ is measurable, $\bm{\xi}_1$ itself must be gauge invariant \citep{bergliaffa2004wave}.
The rotational degree of freedom $\bm{\xi}_1$ is perpendicular to the background vorticity $\bm{\zeta}_0$, so that $\bm{\xi}_1\cdot\bm{\zeta}_0=0$, which can be deduced from \eqref{vort} and \eqref{gauge_inv}.

As alluded to above, one advantage of writing the action \eqref{action_pert} for the perturbations is that it is straightforward to identify conserved quantities~\citep{Richartz2025}.
For the purpose of our analysis, it is convenient to convert to complex fields for the perturbations in the action (e.g. $h_1^2\to |h_1|^2$ etc.) which one can always do since the equations of motion are linear and the physical fields correspond to the real part. Then, the action \eqref{action_pert} is invariant under continuous phase rotations of the form $X_1\to X_1e^{-i\lambda}$, where $X_1$ is the quadruplet of perturbative fields \eqref{quadruplet} and $\lambda$ is an infinitesimal parameter. Then, Noether's theorem implies a conservation law 
\begin{equation}
\partial_t \mathcal{E} + \nabla \cdot \mathbf{J} = 0,     
\end{equation} where
\begin{subequations}
\label{energy}
\begin{align} \label{eps}
    \mathcal{E} = & \ \frac{i}{2}\left(h_1^*\psi_1-h_1\psi_1^*\right) + \frac{ih_0}{2\zeta_0}\hat{\mathbf{e}}_z\cdot\left(\bm{\xi}^*_1\times\bm{\xi}_1\right), \\ \label{J}
    \mathbf{J} = & \ \mathbf{v}_0 \mathcal{E} + \frac{ih_0}{2}\left(\psi_1^*\mathbf{v}_1-\psi_1\mathbf{v}^*_1 \right),
\end{align}
\end{subequations}
which we have written explicitly in terms of gauge invariant quantities. 

\subsection{Perturbations of a non-draining vortex flow}
\label{ssec:lagr-perturbations}

\begin{figure}
\centering
\includegraphics[scale=1]{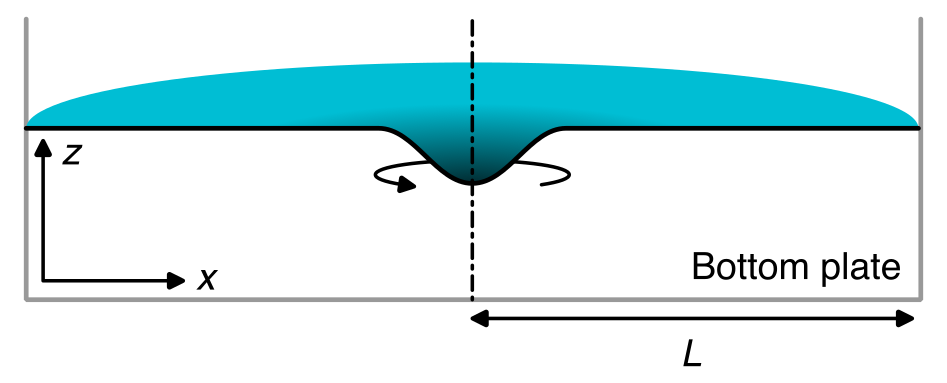}
\caption{Cross section of the non-draining vortex in the $(x,z)$ plane to which we apply our inviscid model. The domain is axially symmetric and, in some cases, spatially constrained by a solid boundary at $r = L$.} \label{fig:2}
\end{figure}

To investigate the effects of a finite-size rotational core and isolate vorticity effects from other factors, we consider an axisymmetric and stationary non-draining vortex, as illustrated in Fig.~\ref{fig:2}. Adopting cylindrical coordinates $(t,r,\theta,z)$ and assuming the shallow water approximation described in \S~\ref{sec:lagrangian}, the background flow field is
\begin{equation} \label{eq:vtheta_generic}
\mathbf{v}_0 = v_\theta(r)\hat{\mathbf{e}}_\theta. 
\end{equation}
The $z$-component of \eqref{euler} gives $P = P_0+\rho g[h_0(r)-z]$ where $P_0$ is an unimportant constant, and $z=h_0$ is the location of the liquid's free surface. Inserting this $P$ into the $r$-component of \eqref{euler} gives $g\partial_r h_0 = v_\theta^2/r$, which is easily integrated to give an analytic expression for the free surface,
\begin{equation} \label{eq:free_surf}
    h_0(r) = h_\infty - \int^\infty_r \frac{v^2_\theta(r') dr'}{gr'}.
\end{equation}
We remark that studying systems whose radial velocity $v_r$ vanishes can be relevant even for draining vortices, as multiple experiments indicate that most of the draining may occur in a thin layer at the bottom of the container, while perturbations at the surface primarily experience an angular velocity field~\citep{andersen2003anatomy,svancara2024rotating,smaniotto2025}.
Regardless, in the same way that superradiant instabilities occur in draining and non-draining potential flows alike, we also expect that generic features of the coupling between potential and rotational perturbations in non-draining flows with vorticity will persist in the draining case.

To study perturbations of a non-draining vortex flow with a finite-size rotational core, characterized by \eqref{eq:vtheta_generic} and \eqref{eq:free_surf}, we apply the framework introduced in \S~\ref{sec:lagrangian}.
Due to the stationarity and the axisymmetry of the flow, perturbations can be decomposed into separate frequency ($\omega$) and azimuthal ($m$) modes according to,
\begin{equation}
    \begin{pmatrix}
        \psi_1 \\ h_1 \\ \bm{\xi}_1
    \end{pmatrix} = \sum_{\omega m} e^{im\theta-i\omega t} \begin{bmatrix}
        \psi_{1\omega m}(r) \\ h_{1\omega m}(r) \\ \bm{\xi}_{1\omega m}(r)
    \end{bmatrix},
\end{equation}
with each mode evolving independently of the others.
To simplify notation, we assume throughout the paper a particular $\omega m$-combination and suppress the associated subscripts. 

Since there is no flow in the radial direction, \eqref{xi_eom} becomes an algebraic equation for $\bm{\xi}_1$, with the particular solution~\citep{richartz2009generalized},
\begin{equation} \label{xi_sol}
    \bm{\xi}_{1} = \frac{\zeta_0}{\Omega^2K} \left[\frac{\hat{\mathbf{e}}_r}{r}\left(-m\Omega+2v_\theta\partial_r\right) \right. \left. +i\hat{\mathbf{e}}_\theta\left(\frac{m\zeta_0}{r}-\Omega\partial_r\right)\right]\psi_{1},
\end{equation}
where we have defined,
\begin{equation} \label{eq24}
    \Omega = \omega-\frac{mv_\theta}{r}, \quad K = 1-\frac{2v_\theta\zeta_0}{r\Omega^2}.
\end{equation}
Using this solution, \eqref{h1_eom} and \eqref{psi1_eom} can be combined into a single differential equation for the $\psi_1$ field,
\begin{equation} \label{1d_eqn}
    \frac{1}{r}\partial_r\left(\frac{c^2r}{K}\partial_r\psi_{1}\right) - V \psi_{1} = 0,
\end{equation}
where the (local) propagation speed of shallow water waves is \citep{torres2018waves}, 
\begin{equation}
    c(r)=\sqrt{gh_0(r)},
\end{equation}
and $V$ is an effective potential barrier given by,
\begin{equation} \label{potential}
    V(r) = -\Omega^2 + \frac{c^2m^2}{r^2}\left(1+\frac{\zeta_0^2}{\Omega^2K}\right) + \frac{m}{r}\partial_r\left(\frac{c^2\zeta_0}{\Omega K}\right).
\end{equation}

\section{Rankine vortex}
\label{sec:linear_stability}

We now specify a particular setup for the background velocity \eqref{eq:vtheta_generic}. We adopt the Rankine vortex model, in which the core rotates as a solid body, while the external flow is purely potential.
These features capture, to some extent, the velocity profile of a realistic vortex and, in the absence of viscosity, any $\mathbf{v}$ of the form \eqref{eq:vtheta_generic} is a solution of the Euler equations when balanced by the free surface profile in \eqref{eq:free_surf}.
The associated velocity field is defined by,
\begin{equation} \label{rankine_vel}
    v_\theta = \begin{cases}
        Cr/R^2, \quad r<R \\
        C/r, \qquad \ r\geq R
    \end{cases},
\end{equation}
where $R$ is the radius of the core, $C$ is the circulation in the irrotational region and we restrict to $C\geq0$ without loss of generality. The solid body core rotates with an angular frequency $C/R^2$. 
Note that the solid body core would normally expand under the influence of viscosity (e.g. the Lamb-Oseen vortex) and a radial flow is required to counteract the expansion and maintain a steady flow (e.g. Burger's vortex~\citep{lautrup2011physics}; see also Appendix \ref{app:viscosity}). Neglecting viscosity therefore amounts to assuming that the expansion or radial flow are weak enough that they do not impact the instabilities we study at leading order.

We consider both open systems, which extend arbitrarily far away from the vortex core, and closed systems, which possess a reflecting boundary at radius $L$ outside the vortex core (cf. Fig.~\ref{fig:2}).
Since we neglect viscosity, we do not have the freedom to impose a no-slip boundary condition, meaning a non-zero angular velocity on $r=L$ is permissible within our approach.
The free surface profile $z=h_0(r)$ of the background flow, obtained by substituting the velocity field \eqref{rankine_vel} into \eqref{eq:free_surf}, is given by 
\begin{equation}
    h_0 = \begin{cases}
        h_c + C^2r^2/(2gR^4), \quad r<R \\
        h_\infty - C^2/(2gr^2), \quad \ \ \,  r\geq R
    \end{cases},
\end{equation}
where $h_c$ and $h_\infty$ are respectively the height of the free surface in the centre, i.e. in the limit $r \rightarrow 0$ and the height in the asymptotic limit $r \rightarrow \infty$.
Continuity at the boundary of the vortex core requires that 
\begin{equation}
    h_c = h_\infty - \frac{C^2}{gR^2}.
\end{equation}
It is convenient to perform the following rescaling,
\begin{equation} \label{adim}
    \frac{r}{R}\to r, \quad \frac{t\sqrt{gh_\infty}}{R}\to t \quad \frac{v_\theta}{\sqrt{gh_\infty}}\to v_\theta, \quad \frac{h}{h_\infty}\to h, \quad \frac{C}{R\sqrt{gh_\infty}} \to C.
\end{equation}%
In these units, both the asymptotic wave speed $\sqrt{g h_\infty}$ and the radius of the vortex core are unity. Consequently, the problem depends only on the dimensionless circulation and, in the case of the closed system, also on the dimensionless ratio $L/R$.
Note that these units are equivalent to setting $g=h_\infty=R=1$ in the original units.

\begin{figure}
\centering
\includegraphics[scale=0.8]{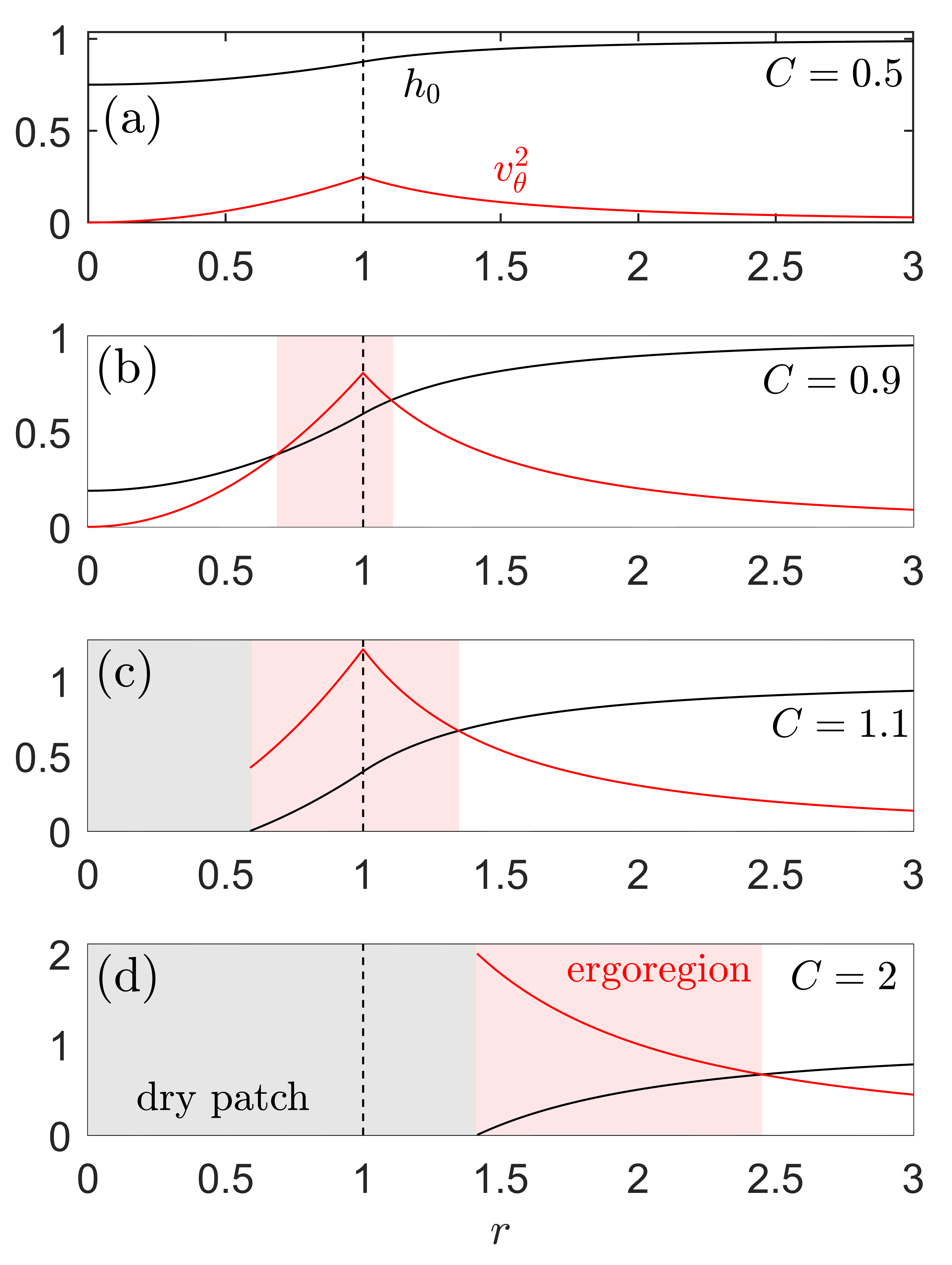}
\caption{Background flow field solutions for four different $C$ values in non-dimensional units, see Eq.~\eqref{adim}. The black (red) curve is $h$ ($v_\theta^2$) as a function of $r$.
The dashed black line is the boundary of the rotational core of the vortex. The pink shaded region represents the ergoregion where $v_\theta^2 > h$ (the flow exceeds the wave speed), whilst the grey shaded region represents the dry patch.
(a): at low $C$, there is no ergoregion and the rotational core covers the whole central region. (b): as $C$ increases, an ergoregion develops around the edge of the rotational core. (c): increasing $C$ further, a dry patch develops in the centre. (d): for high enough $C$, the boundary of the dry patch is in the potential region ($r > 1$) and the rotational core is absent from the flow.} \label{fig:3}
\end{figure}

The Rankine vortex exhibits three qualitatively distinct regimes depending on the value of $C$. The various cases are illustrated in Fig.~\ref{fig:3} and described below.
\begin{enumerate}
    \item~\emph{Wet-plate} (WP): the fluid covers the entire region $r\geq 0$ in the case of an open system and $0 \le r \le L$ in the case of a closed system. This regime occurs for (the dimensonless) $C<1$.
    \item~\emph{Dry-plate rotational} (DPR): the free surface intersects the bottom plate of the container at \mbox{$r_\dpr = \sqrt{2(1-1/C^2)}$,} inside the rotational core. Hence, there is a dry patch in the region $0 \le r<r_\dpr$ which is excluded from the fluid domain. The fluid covers the region $ r \geq r_\dpr$ in the case of an open system and $r_\dpr \le r \le L$ in the case of a closed system. This regime occurs for $1\leq C<\sqrt{2}$.
    \item~\emph{Dry-plate potential} (DPP): the free surface intersects the bottom plate at \mbox{$r_\dpp = C/\sqrt{2} > R$,} i.e. in the region of potential flow. Hence, the flow is everywhere irrotational. The fluid covers the region $ r \geq r_\dpp$ in the case of an open system and $r_\dpp \le r \le L$ in the case of a closed system. This regime occurs for $C\geq\sqrt{2}$.
\end{enumerate}%
As we discuss in \S~\ref{sec:intro}, region where \mbox{$|v_\theta(r)|>\sqrt{gh_0(r)}$} marks the effective ergoregion. Within our model, its presence relates to the existence of negative-energy waves in the system, i.e. negativity of the energy density $\mathcal{E}$, Eq.~\eqref{eps}. In the WP regime, an annular ergoregion forms when $\sqrt{2/3}<C\leq 1$, covering the range \mbox{$r_{e}^- \le r \le r_{e}^+$}, where \mbox{$r_{e}^- = \sqrt{2(1/C^2-1)}$} and \mbox{$r_{e}^+=C\sqrt{3/2}$}. In the DPR and DPP regimes, an ergoregion is always present, extending over the range \mbox{$r_\dpr \le r \le r_{e}^+$} and \mbox{$r_\dpp \le r \le r_{e}^+$}, respectively (see Fig.~\ref{fig:3}).

\subsection{Boundary conditions} \label{sec:boundarycs}

We need to specify boundary conditions on both ends of the fluid domain to solve the wave equation \eqref{1d_eqn}. Recall that the inner boundary corresponds either to the centre of the container ($r = 0$) in the WP regime, or to the point where the free surface meets the bottom plate, i.e.~at $r = r_\dpr$ in the DPR regime and $r = r_\dpp$ in the DPP regime. In all three regimes, the outer boundary corresponds either to the position of the reflective wall in the closed problem at $r = L$, or to the asymptotic limit $r \to \infty$ in the open problem. At the inner boundary, we require that the perturbation field $\psi_1$ is ``well-behaved'', so that the flow perturbations do not diverge as functions of $r$. At the outer boundary, in the closed problem we assume that the perturbation of the radial velocity component must vanish while in the open problem we assume that there is no wave incoming from infinity.

\subsubsection{Inner boundary condition}

The explicit form of the inner boundary condition depends on the specific regime under consideration, as explained above. In the WP and DPR regimes, we use the fact that inside the rotational core ($r < 1$), the wave equation~\eqref{1d_eqn} can be recast as a hypergeometric differential equation~\citep{Florentin1966}. We perform a coordinate and field transformation, $z = (r / r_\dpr)^2$ and $\psi_1(z) = z^{m/2}\phi(z)$, where $r_\dpr$ serves as a convenient rescaling parameter with a direct physical meaning only in the DPR case, and $m$ is the azimuthal number. Using the proportionality $v_\theta \propto r$ in this region, the wave equation~\eqref{1d_eqn} then takes the form,
\begin{equation}
\label{hypergeo}
z(1-z)\frac{d^2\phi}{dz^2} + [1+m - (m+2)z]\frac{d\phi}{dz} - \left(\frac{2m +Q}{4}\right)\phi=0,
\end{equation}
where the parameter $Q$ is defined as,
\begin{equation}
    Q = 2 \left[\frac{2}{1-\omega/ m C}+m^2 \left(1-\frac{\omega}{m C}\right)^2-4\right].
\end{equation}
In terms of the new coordinate $z$, the inner boundary is located at $z=0$ in the WP regime and at $z=1$ in the DPR regime. We note that both $z=0$ and $z=1$ are singular points of the hypergeometric equation above. We must therefore determine the regular solutions of \eqref{hypergeo} around these singular points. In the WP regime, we find that the regular solution valid inside the rotational core $0 \le r \le 1$ is
\begin{equation} \label{psi1_wp}
  \psi_1 \propto r^{m} \, {}_2F_1\left[\alpha_+,\alpha_-,1+m,(r/r_\dpr)^2\right],
\end{equation}
where ${}_2F_1$ is the Gaussian hypergeometric function with coefficients $\alpha_\pm$ given by
\begin{equation}
   \alpha_\pm = \frac{1+m\pm\sqrt{1+m^2-Q}}{2}.
\end{equation} 
In the DPR case, we find that the regular solution valid inside the rotational core $r_\dpr \le r \le 1$ is
\begin{equation} \label{psi1_dpr}
\psi_1(r) \propto r^{m} \,{}_2F_1\left[\alpha_+,\alpha_-,1,1-(r/r_\dpr)^2\right].
\end{equation}

Finally, since the inner boundary in the DPP regime is located where the flow is irrotational, we cannot use the hypergeometric equation anymore. Instead, we use the Frobenius method to determine a pair of linearly independent series solutions of the wave equation \eqref{1d_eqn}. We find that, around $r=r_\dpp$, the regular solution of the wave equation is given by 
\begin{equation} \label{frob}
    \psi_1 \propto 1 - \left(\omega- \frac{mC}{r_\dpp^2}   \right) ^2 \frac{r_\dpp ^3}{C^2}(r-r_\dpp) + \mathcal{O}(r-r_\dpp)^2.
\end{equation}
From this we deduce that the appropriate inner boundary condition in the DPP is a relation between the field and its derivative of the form,
\begin{equation} \label{bc_inner_dpp}
    \frac{\psi_1 '(r_\dpp)}{\psi_1 (r_\dpp)} =  - \left(\omega- \frac{mC}{r_\dpp^2}   \right) ^2 \frac{r_\dpp ^3}{C^2}.
\end{equation}

\subsubsection{Outer boundary condition}

Explicitly, the outer boundary condition in the closed problem is
\begin{equation} \label{bc_outer1}
 \psi_1'(L)=0,   
\end{equation}
so that $\psi_1$ satisfies a Neumann boundary condition at the reflective surface. In the open problem, the explicit form of the outer boundary condition follows from the asymptotic solution of \eqref{1d_eqn}, given by 
\begin{equation} \label{asymp}
    \psi_1 \approx \frac{A^+_\infty e^{i\omega r} +  A^-_\infty e^{-i\omega r}}{\sqrt{r}} + \mathcal{O} (r^{-3/2}),
\end{equation}
where $A^{\pm}_\infty$ are the amplitudes of the radially out-going and in-going waves (specifically, the adiabatic part of the amplitude which does not depend on $r$). The above relation can be inverted to express the wave amplitudes in terms of $\psi_1$ and its derivative, yielding
\begin{equation} \label{wave_amps}
    A^\pm_\infty = \lim_{r\to\infty} \left[\left(i\omega\pm\frac{1}{2r}\right)\psi_1(r) \pm\psi_1'(r)\right]\frac{\sqrt{r}e^{\mp i\omega r}}{2i\omega}.
\end{equation}
The requirement of no incoming waves corresponds to $A^-_\infty = 0$, which, in numerical calculations, must be implemented at some sufficiently large radius $r_\far \gg 1$. Hence, from \eqref{wave_amps}, follows the mixed boundary condition
\begin{equation} \label{bc_outer2}
     \left(i \omega-\frac{1}{2r_\far} \right)\psi_1 (r_\far) - \psi_1'(r_\far) = 0. 
\end{equation}
In practice, we take $r_\far=10\,\mathrm{max}(1,r_\tp^+(\omega),2\pi\mathrm{Re}[\omega]^{-1})$, where $r_\tp^+$ is defined in Appendix~\ref{app:wkb} and $\omega$ is our initial guess for the frequency that gets updated during the numerical algorithm (see \S~\ref{sec:num_int}).

\subsection{Matching conditions at the rotational core interface}

In the WP and DPR regimes, one must account for the discontinuity in the derivative of the velocity field at the surface $r=R$, which separates the rotational core from the irrotational part of the flow.
In other words, in addition to the boundary conditions discussed in \S~\ref{sec:boundarycs}, we also need to match the wave fields of the rotational core, given either by \eqref{psi1_wp} or \eqref{psi1_dpr}, with the wave field of the irrotational regime at $r=1$.
To connect our two solutions, we integrate \eqref{1d_eqn} in an infinitesimal region around $r=1$ to find the following matching conditions, 
\begin{subequations} \label{matching} 
\begin{align}   
    \psi_1(1^+)= & \ \psi_1(1^-), \\
    \psi'_1(1^+) = & \ \frac{1}{K_c}\left[\psi'_1(1^-) -\frac{m\zeta_c}{\Omega_c}\psi_1(1^-)\right],
\end{align}
\end{subequations}
with $1^+$ and $1^-$ indicating, respectively, the limit as $r\to 1$ from above and from below, and subscript $c$ denotes the constant ($r$-independent) values of these quantities inside the rotational core, i.e.
\begin{equation}
    \Omega_c = \omega-mC, \quad \zeta_c = 2C, \quad K_c = 1-\frac{\zeta_c^2}{\Omega_c^2}.
\end{equation}
These equations lay the foundation for investigating the stability of the Rankine vortex to perturbations inside and outside the rotational core. Findings of our stability analysis are presented in the following section.

\section{Numerical integration}
\label{sec:num_int}

We start by determining complex eigenfrequencies $\omega$ and the associated eigenfunctions $\psi_1$  by solving the eigenvalue problem consisting of Eq.~\eqref{1d_eqn} and the relevant set of boundary conditions. We specifically search for unstable modes  characterised by \mbox{$\mathrm{Im}[\omega]>0$}, i.e. growing in time with a characteristic timescale of \mbox{$\mathrm{Im}[\omega]^{-1}$}. The real part of the frequency \mbox{$\mathrm{Re}[\omega]$} sets the oscillation rate of the unstable mode.

Our main numerical technique involves the direct integration of Eq.~\eqref{1d_eqn}, starting from the inner region and moving to large $r$. In the WP and DPR regimes, we integrate from the boundary of the vortex core ($r=1$) up to $r=L$ (closed problem) or up to $r=r_\far$ (open problem). Using a shooting method, we solve for the eigenfrequencies $\omega$ which satisfy either \eqref{bc_outer1} or \eqref{bc_outer2} and simultaneously \eqref{matching}.
Similarly, in the DPP regime, we integrate \eqref{1d_eqn} from $r_\dpp$ up to the outer boundary. As before, a shooting method is applied to solve for the eigenfrequencies that simultaneously satisfy the inner boundary condition \eqref{bc_inner_dpp} and the outer boundary condition, given by \eqref{bc_outer1} or \eqref{bc_outer2}.

As an example, we first study excitations with $m = 2$ (waves with two crests and troughs in the azimuthal direction) in \S~\ref{sec:mequals2}, before studying higher $m$-modes in \S~\ref{sec:m34}. In the latter, we employ a separate numerical model called the continued fraction method (CFM; see Appendix~\ref{app:cfm} for details). We report no discrepancies between the two approaches, thereby validating their consistency. Our analysis is further supported by a Wentzel-Kramers-
Brillouin (WKB) calculation, implemented for the DPP regime in Appendix~\ref{app:wkb}.

\subsection[Eigenmodes for m=2]{Eigenmodes for $m=2$} 
\label{sec:mequals2}
\subsubsection{Open system} \label{sec:open}

\begin{figure}
\centering
\includegraphics[scale=0.8]{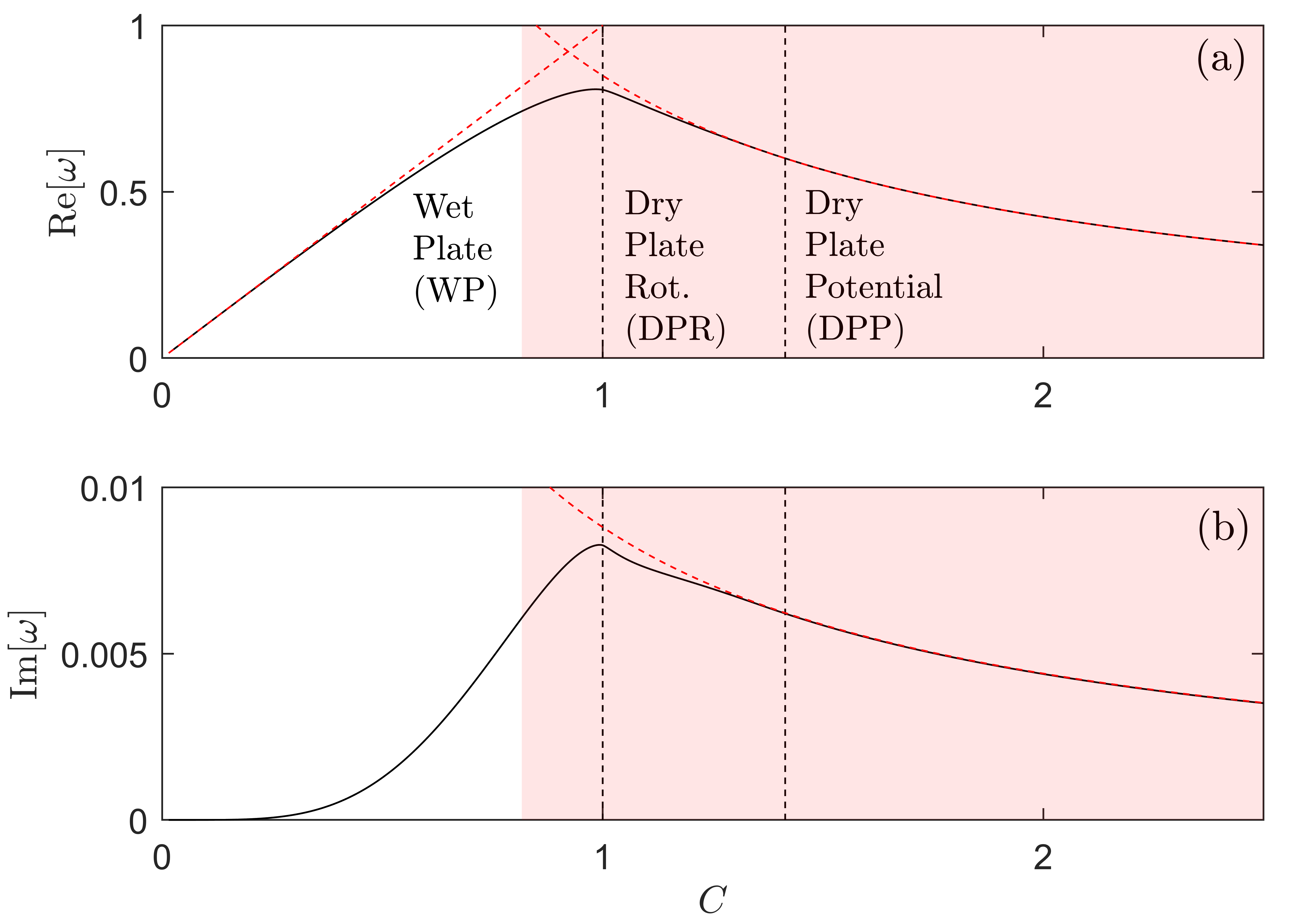}
\caption{Eigenfrequencies of the unstable $m=2$ mode (solid black lines) as a function of $C$ for the open system. The oscillation rate \mbox{$\mathrm{Re}[\omega]$} is shown in panel (a), while the growth rate \mbox{$\mathrm{Im}[\omega]$} is presented in panel (b). The dashed vertical lines indicate the boundaries of different flow regimes. At low $C$, the flow is wet-plate (WP). An ergoregion is present everywhere inside the pink region. 
As we increase $C$, the flow becomes dry-plate rotational (DPR) at the first black dashed line and dry-plate potential (DPP) at the second one. The dashed red lines indicate the $1/C$ fall off at large $C$ and the linear dependence at low $C$ in the real part.} \label{fig:4}
\end{figure}

As outlined in \S~\ref{sec:intro}, we expect an instability similar to the ergoregion instability in the large $C$ limit, since there is no mechanism to prevent waves from carrying positive energy to infinity. We would like to understand what happens as $C\to0$ and the vortex develops a rotational core. Note in passing that we focus on $m=2$ waves since our numerical investigation demonstrates that the ergoregion instability does not occur for $m=1$ waves in the DPP regime. In brief, there are no excitations with sufficiently short wavelength to fit between the edge of the dry patch and the peak of the effective potential $V$ defined by Eq.~\eqref{potential}.

Our results are presented in Fig.~\ref{fig:4}, where we plot the real and imaginary components of $\omega$ as a function of $C$ (black solid lines). We find that the open system is unstable for all values of $C$ due to the presence of an eigenmode with $\mathrm{Im}[\omega]>0$. However, for large $C$, we find that the frequency of this eigenmode falls off as $1/C$ (red dashed line). An approximate WKB treatment, presented in Appendix~\ref{app:wkb}, explicitly demonstrates that negative energy inside the ergoregion underpins the existence of this unstable mode, allowing us to identify it with the ergoregion instability. The unstable mode consists of a negative energy oscillation inside the analogue ergoregion (cf. Fig.~\ref{fig:3}(d)) which couples to a positive energy excitation radiating to infinity.

Let us follow the frequency of the unstable mode in Fig.~\ref{fig:4} as $C$ decreases from large to small values. At the transition from the DPP regime to the DPR regime, $\mathrm{Re}[\omega]$ begins to deviate from the $1/C$ fall-off which is characteristic of the ergoregion instability in an irrotational flow. In the WP regime, we find that $\mathrm{Re}[\omega]\to 0$ as $C\to 0$. In particular, the oscillation rate in the limit of vanishing $C$ decreases linearly according to   
\begin{equation}  
\mathrm{Re}[\omega ] =(m-1)C + \mathcal{O}(C^3). 
\end{equation}
This behaviour, markedly different from that of the irrotational flow, is an indicator of the additional physics involved in flows with vorticity. From the analogue gravity perspective, one typically cares about instabilities associated with the ergoregion and not the vorticity field, hence, it is crucial to understand the transition between these separate regimes.
From the fluid dynamical perspective, this transition is associated with the coupling of different families of waves, as we now show.

\subsubsection{Closed system (small container)}

To understand the origin of the instability at low $C$ observed in the open system, we now study the closed system. The advantage with the latter is that energy is retained by the system because of the reflective barrier at $r=L$, and the equations of motion become  conservative. In that case, energy is a conserved quantity, allowing us to easily analyse the energy content of the different eigenmodes.

Additionally, in a closed system, the spectrum of modes outside the vortex is discretised by the reflective barrier. These modes are positive-energy standing waves on the free surface and, in the static limit ($C = 0$), they are Bessel functions. If the container is sufficiently small, the lowest-frequency surface wave will have a higher frequency than the negative-energy vortex mode, hence the system remains stable. In other words, the surface waves and the vortex mode can be identified as different eigenmodes with distinct real eigenfrequencies $\omega$.

\begin{figure}
\centering
\includegraphics[scale=0.8]{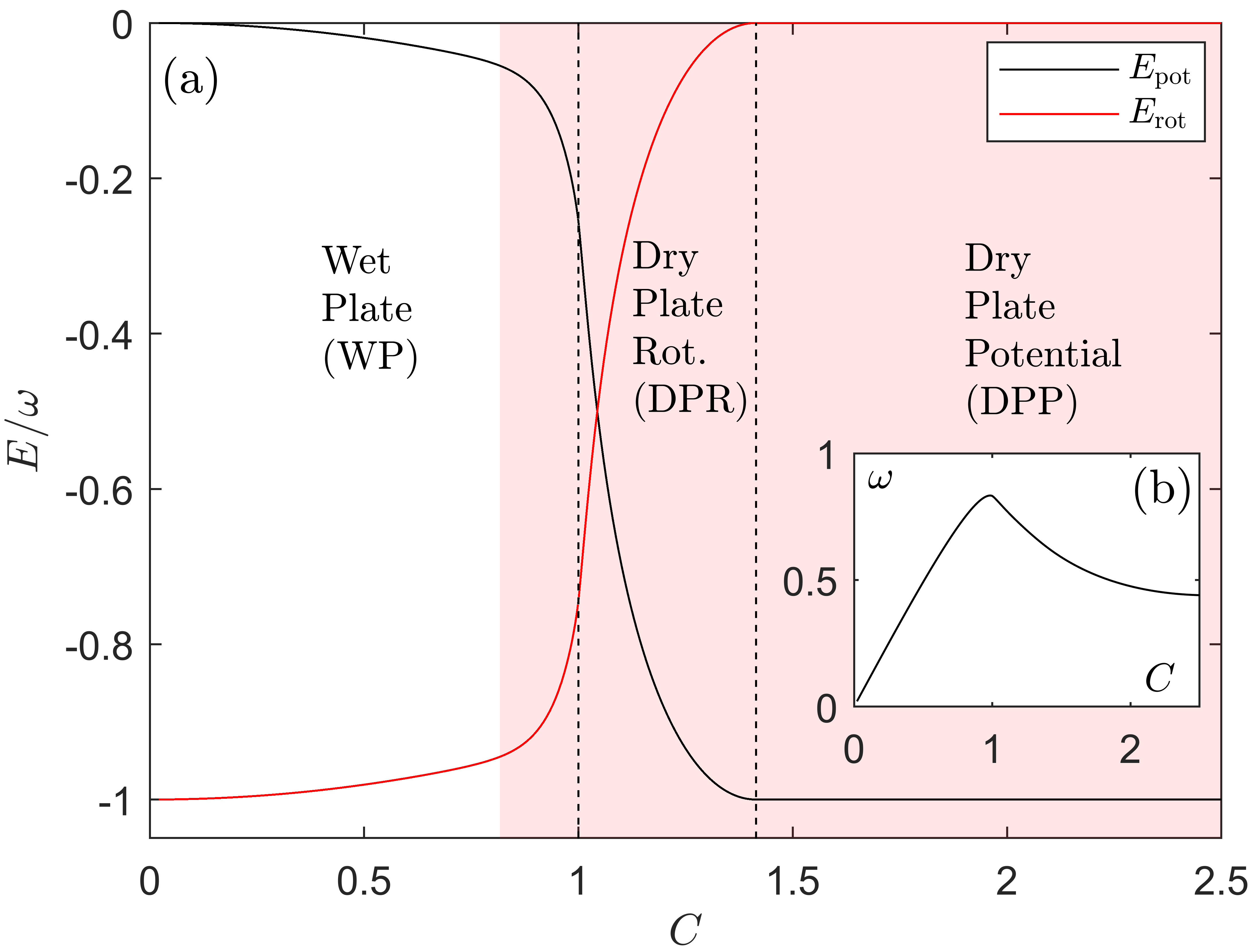}
\caption{Energy components of the $E<0$ mode with $m=2$ for $L=3.5$. In the WP regime, $E_\mathrm{rot}$ is the dominant energy component and the eigenmode is due to an excitation of the vorticity field. In the DPP regime, the flow is potential and the negative energy of this state is the result of the ergoregion. The presence of an ergoregion is indicated by the shaded pink area. The inset (b) shows the mode frequency, which resembles $\mathrm{Re}[\omega]$ in the case of the open system [see Fig.~\ref{fig:4}(a)], with the deviations from the expected $1/C$ behaviour at large $C$ resulting from finite size effects.} \label{fig:5}
\end{figure}

The energy $E$ of each eigenmode is defined by 
\begin{equation}
 E = \omega\int d^2\mathbf{x} \, \mathcal{E},   
\end{equation}
with $\mathcal{E}$ given in \eqref{energy}. We assume the normalization $\left|\int d^2\mathbf{x} \, \mathcal{E}\right|=1$, and split the energy into potential and rotational components,
\begin{equation}
\begin{split}
    E_\mathrm{pot} = & \ \omega\int d^2\mathbf{x}\, \mathrm{Im}[h_1\psi_1^*], \\
    E_\mathrm{rot} = & \ \omega\int d^2\mathbf{x} \, \frac{h_0}{\zeta_0} \, \mathrm{Im}[\xi_{1r}\xi_{1\theta}^*].
\end{split}
\end{equation}
Using the notion of energy introduced above, we find that some of the eigenmodes are characterised by $E>0$ and others by $E<0$. In Fig.~\ref{fig:5}, we display the energy components of the \mbox{$m=2$} eigenmode with $E<0$ for a small container of size $L=3.5$. 
Since $E_\mathrm{rot}\propto\zeta_0$, this component vanishes in the large $C$ (DPP) regime, and all negative energy is due to $E_\mathrm{pot}$.
As in the open system, negativity of $E_\mathrm{pot}$ is due to the existence of an ergoregion. 
The contribution of $E_\mathrm{pot}$ to the total energy diminishes as $C$ is lowered until $E_\mathrm{rot}$ is the dominant contribution.
The transition begins when the vortex develops a rotational core, with $E_\mathrm{pot}$ and $E_\mathrm{rot}$ becoming comparable near the transition to the WP regime. When the ergoregion is absent, all the mode energy is essentially in the $\bm{\xi}$ field, characterising the eigenmode as an oscillation of the vorticity field.

To illustrate the difference between rotational and irrotational
degrees of freedom, we study the spatial profile of the total height ($h=h_0+\epsilon h_1$) and vorticity ($\zeta=\zeta_0+ \epsilon\zeta_1$) fields associated with the $E<0$ eigenmode, where $\epsilon$ is a small amplitude. The height variations are given by inverting \eqref{psi1_eom} for $h_1$. Conversely, using Eq.~\eqref{1d_eqn} to substitute for a second spatial derivative, we find that the vorticity perturbations are,
\begin{equation}
    \zeta_1 = \hat{\mathbf{e}}_z\cdot(\grad\times\bm{\xi}_1) = \frac{i\zeta_0}{\Omega_c K_c h_0}\left(h_0'\psi_1'+ \Omega_c^2\psi_1\right),
\end{equation}
with primes denoting radial derivatives. This expression is manifestly non-zero only inside the rotational core.
Although it looks like it diverges in the DPR regime when $h_0\to 0$, we find by expanding $\psi_1$ in powers of $(r-r_\dpr)$ that,
\begin{equation}
    \zeta_1 = \frac{i\zeta_c}{2\Omega_c K_c}\left(\frac{m^2}{r_\dpr^2}-\frac{\Omega_c^2}{h'_\dpr}\right) + \mathcal{O}(r-r_\dpr),
\end{equation}
where $h'_\dpr=h'_0(r_\dpr)$. Hence, $\zeta_1$ is finite when $r \to r_\dpr$. After restoring the $t$ and $\theta$ dependence, we eventually find that
\begin{equation}
    \begin{pmatrix}
        h \\ \zeta
    \end{pmatrix} = \begin{pmatrix}
        h_0 \\ \zeta_0
    \end{pmatrix} - \begin{pmatrix}
        \Omega\psi_1 \\ \mathrm{Im}[\zeta_1]
    \end{pmatrix}\epsilon\sin(m\theta-\omega t).
\end{equation}

\begin{figure}
\centering
\includegraphics[scale=0.8]{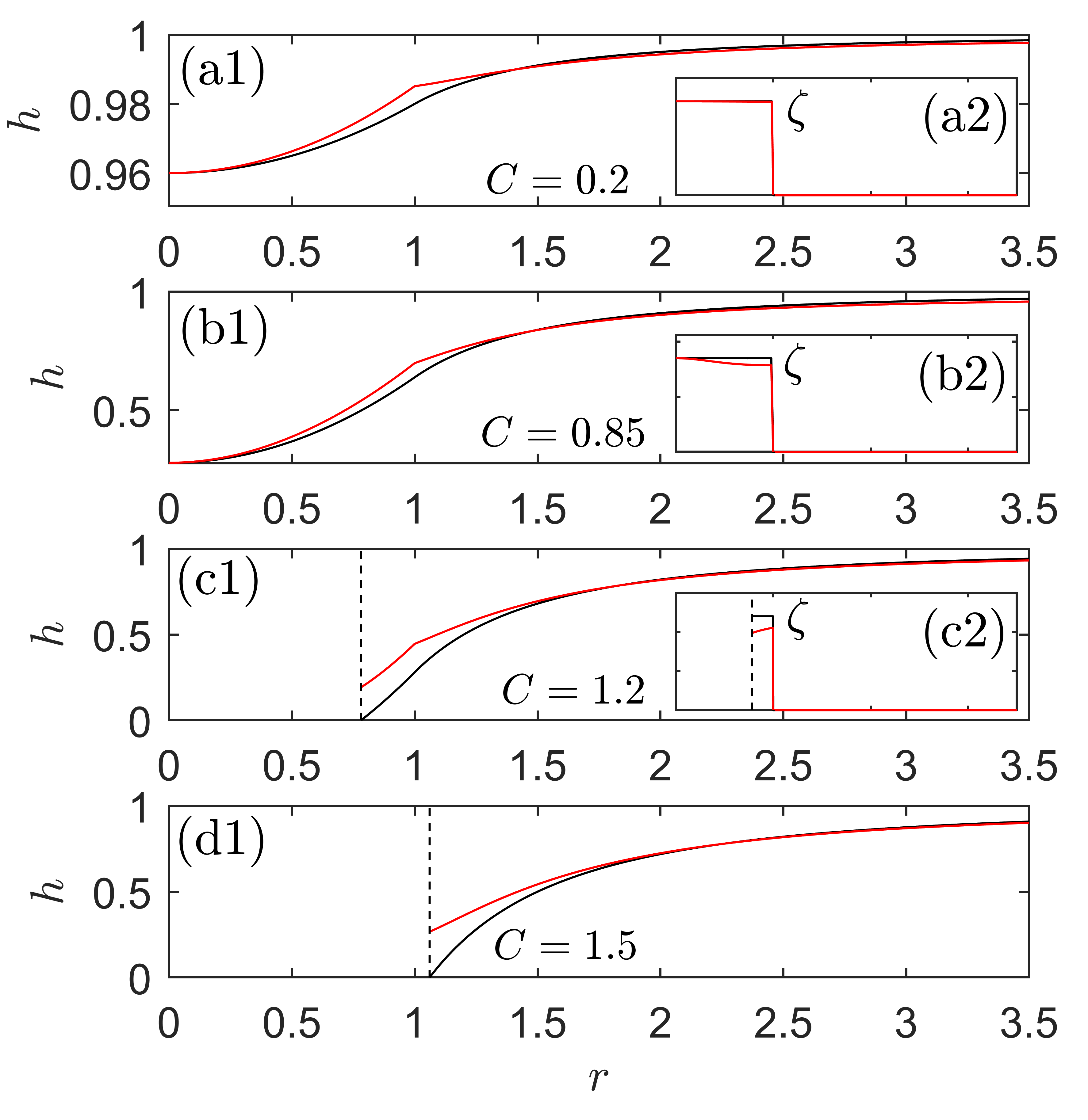}
\caption{Cross-sections of the stationary height profiles (black curves) and full height profiles including the unstable mode (red curves) with $m=2$, $L=3.5$ and $\epsilon=0.2$. The insets show the vorticity profile. In the WP regime (top two panels), the instability is associated with an oscillation of the edge of the vorticity patch and the height profile is pinned in the centre. In the DPR regime (third panel), the entire vorticity patch oscillates and $h$ is not pinned at the inner boundary. This translates physically to a displacement of the boundary of the dry patch responsible for the rotating polygons instability.
Note that the discontinuities around $r=1$ are a feature of the Rankine vortex \eqref{rankine_vel} and would be smoothed over by a more realistic velocity profile.
In the DPP regime (fourth panel), there are no vorticity oscillations and only $h$ oscillates. } \label{fig:6}
\end{figure}

In Fig.~\ref{fig:6}, we illustrate a cross-section of these profiles (red lines) at the maximum of the sine term. Moving round in the $\theta$ direction, variations about $(h_0,\zeta_0)$ oscillate with periodicity $2\pi/m$ and, under time evolution, orbit the vortex in the $\mathrm{sgn}(\omega/m)$ direction. In the WP regime, the $E<0$ mode is an oscillation of the edge of the rotational core, i.e. a Kelvin-Kirchoff wave \citep{fabre2014generation}. This is apparent from the deviation of $h$ away from its mean value around $r=1$ in panels (a1) and (b1). Furthermore, panel (b2) clearly shows a deviation away from the constant vorticity at the edge of the rotational core. Notice in particular how, in the WP regime, the perturbation goes to zero at the centre of the vortex. By contrast, in the DPR regime shown in panel (c1), the free surface shows a strong deformation all the way up to the edge of the dry patch in the centre. Correspondingly, the entire vorticity field oscillates as shown in panel (c2). Finally, in the DPP regime, the rotational core is completely absent and the $E<0$ mode corresponds to a surface deformation near the centre.

\subsubsection{Closed system (large container)}

\begin{figure}
    \centering
    \includegraphics[width=1\linewidth]{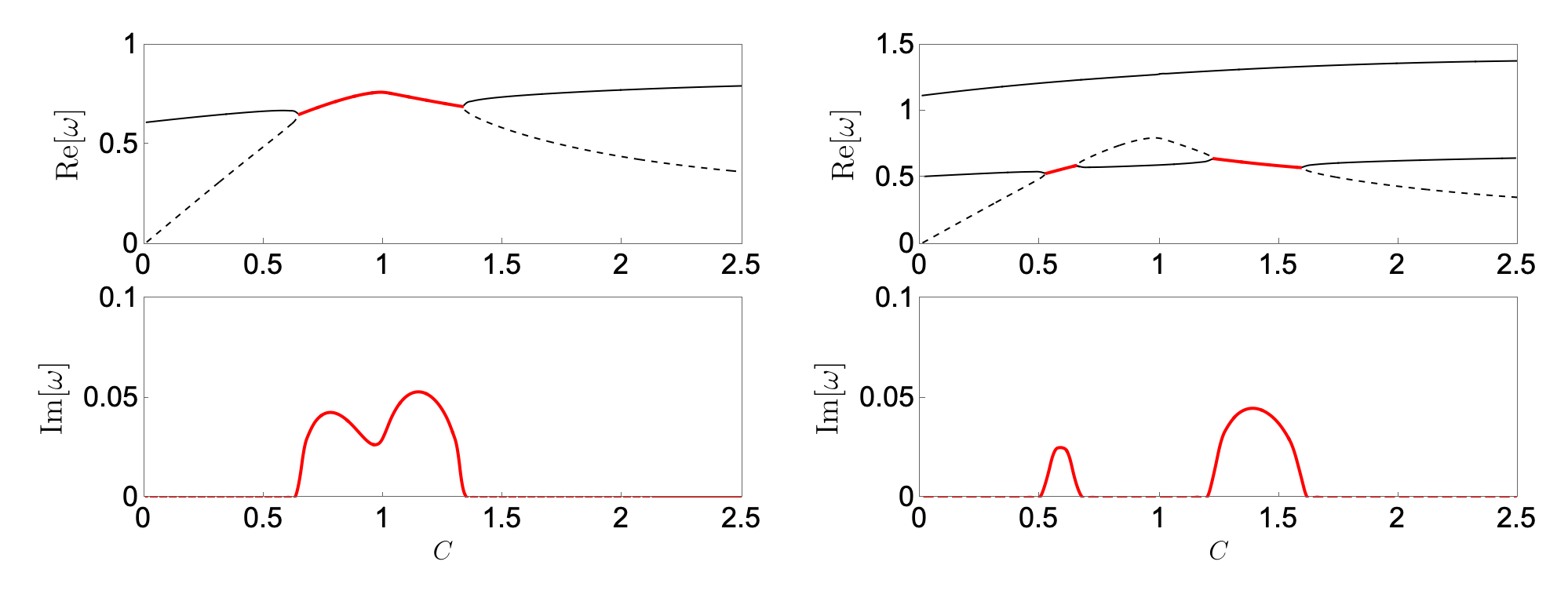}
    \caption{Eigenfrequencies $\omega$ for the $m=2$ mode of the vortex in a closed system with sizes $L=5$ (left) and $L=6$ (right). The upper panels display $\mathrm{Re}[\omega]$ while the bottom panels show $\mathrm{Im}[\omega]$. Solid black lines correspond to surface waves and dashed black lines correspond to vortex modes. When the frequencies of a surface wave mode and a vortex mode coalesce, the two modes acquire an imaginary part and the vortex becomes unstable. The unstable mode, i.e. growing in time, is shown in red. It always appears alongside a complex conjugate mode, with equal and opposite-signed $\mathrm{Im}[\omega]$ that decays in time.} \label{fig:7}
\end{figure}

As discussed in the previous section, if the container is finite (closed system) and sufficiently compact, the vortex is stable since there is no dissipation mechanism to excite negative energy modes. However, since there is a discrete spectrum of positive energy surface (standing) waves outside the vortex, as the size $L$ of the container increases the frequencies of the vortex mode and surface waves become comparable, and the two combine to produce a dynamical instability. Specifically, the two modes split into a complex conjugate pair of zero energy eigenmodes with complex frequencies $\omega\in\mathbb{C}$. The one with $\mathrm{Im}[\omega]>0$ is the unstable mode whilst the one with $\mathrm{Im}[\omega]<0$ is a consequence of the time reversal symmetry of the problem (note, the open problem is not symmetric under time reversal since then the outgoing boundary condition becomes incoming). This is the situation described by the hybrid instability in Fig.~\ref{fig:1}. Physically, the oscillation of the vortex interior lowers its energy, and this energy is transferred into an oscillation of the surface outside the vortex core. Radiating energy into a standing surface wave is the finite size analogue of what happens in the open system, where the energy is radiated into an outgoing wave.

Since the frequency spacing between surface waves is roughly $\pi/L$, the instability only occurs for narrow intervals of $C$ values (when $L$ is not too large). This situation is shown in Fig.~\ref{fig:7} for containers of size $L=5$ and $L=6$. The solid black lines in the upper panels are the frequencies of surface waves, which are characterised by $\mathrm{Im}[\omega] = 0$ and tend to increase with $C$. These are to be compared with the frequencies of the vortex modes, indicated by dashed black lines.

When the frequency of a surface wave crosses that of the vortex mode, a ``bubble of instability'', corresponding to $\mathrm{Im}[\omega] > 0$, appears in the bottom panels of Fig.~\ref{fig:7}. If we were to increase the size $L$ of the container further, the frequency spacing between the different surface waves would decrease as $1/L$ and more unstable bubbles would appear on the $C$-axis in Fig.~\ref{fig:7}, eventually overlapping for $L$ large enough~\citep[see e.g.][]{giacomelli2020ergoregion}. In the limit $L\to\infty$, the spectrum of surface waves approaches a continuum and one recovers the open problem studied in \S~\ref{sec:open}. In this case, the density of states is sufficiently high that the vortex mode can always interact with some surface wave.

\subsection{Higher $m$ eigenmodes} \label{sec:m34}

\begin{figure} 
    \centering
    \includegraphics[width=1\linewidth]{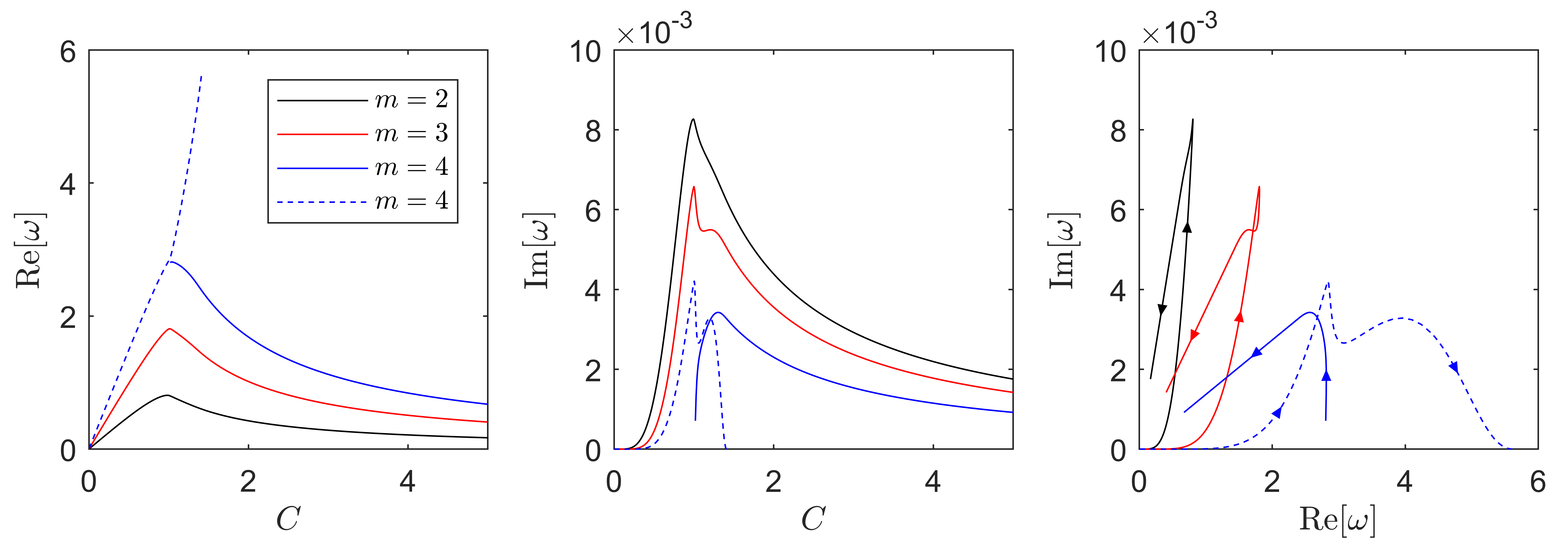}
    \caption{Unstable mode eigenfrequencies for $m=\{2,3,4\}$ (solid coloured lines), calculated using a continued fraction method. We display the variation of $\mathrm{Re}[\omega]$ and $\mathrm{Im}[\omega]$ with $C$ as well as the mode trajectories through the complex frequency plane. In the third panel, the arrows indicate the direction of increasing $C$. For $m=4$, the low-$C$ vorticity instability (dashed blue line) is a distinct mode from the ergoregion instability at large $C$.} \label{fig:8}
\end{figure}

So far we have focused on eigenmodes with $m=2$, the smallest azimuthal number for which we find unstable modes. We expect that the general characteristics observed in \S~\ref{sec:mequals2} for $m=2$ persist for higher azimuthal numbers. To confirm this, we consider the illustrative cases of $m=3$ and $m=4$, comparing the eigenfrequencies with the previous ones for $m=2$ to determine the qualitative and quantitative differences.

In Fig.~\ref{fig:8} we display the complex eigenfrequency of the unstable mode for $m=\{2, 3, 4\}$, obtained through the continued fraction method (see Appendix~\ref{app:cfm}). We remark in passing that the results for $m=2$ are equivalent to those obtained by direct integration and are displayed in Fig.~\ref{fig:4}. Here we observe that $\mathrm{Re}[\omega]$ tends to increase with $m$ whilst $\mathrm{Im}[\omega]$ decreases. The latter is explained by the fact that (in the region where the flow is potential) the amount of superradiant amplification decreases for larger $m$, rendering the growth of higher $m$-modes slower. We also observe that the low and high $C$ behaviour of the eigenfrequencies, discussed in the last paragraph of \S~\ref{sec:open} and highlighted in Fig.~\ref{fig:4} for $m=2$, also applies to $m=\{3, 4\}$.

However, despite the similarities, we also observed qualitative differences. For $m=3$, the transition from vorticity-induced to ergoregion-induced instability is less smooth than in the case of $m=2$, displaying a sudden dip in $\mathrm{Im}[\omega]$ around $C=1.1$. This corresponds to the mode trajectory (parametrised by $C$) forming a loop in the complex plane. For $m=4$, the dip in $\mathrm{Im}[\omega]$ becomes more pronounced and we find that the low $C$ vorticity instability is a distinct mode from the ergoregion one observed for high $C$.

This contrasts with what happens for $m=\{2, 3\}$ where the vorticity instability at low $C$ smoothly transitions into the ergoregion instability at large $C$. Furthermore, there is a range of circulation parameters, specifically $1.02 \lesssim C \lesssim 1.41$, where the two distinct unstable $m=4$ modes coexist (see Fig.~\ref{fig:8} where the solid and dashed blue lines overlap).

\section{Conclusion and outlook}

To summarize, we have studied the stability of a stationary, swirling flow modelled by a Rankine vortex in the long-wavelength limit.
Facilitated by our assumption of an inviscid fluid, we applied a variational framework to investigate the mechanism that underpins surface wave instabilities. In all cases, the instability is underpinned by a negative energy mode which is localised in the vortex core. In the open system, this mode gets excited when the vortex spontaneously radiates surface waves to infinity. When the vortex is placed in an impermeable cylindrical container, the frequency of the negative energy mode has to be matched to one of the surface eigenmodes for an instability to occur.
We then identified two limiting behaviours corresponding to qualitatively distinct destabilising mechanisms. At low circulation, the negative energy mode manifests as a disturbance of the vorticity field near the edge of the rotational core. 
In contrast, at high circulation, potential perturbations acquire negative energy inside the vortex, leading to a deformation of the free surface at the boundary of the central dry patch. 
Building on previous analyses of free-surface vortex instabilities~\citep{tophoj2013rotating,fabre2014generation,mougel2014waves,mougel2017instabilities}, our framework explicitly separates the energetic contributions of irrotational and vorticity perturbations and identifies the regimes in which each type of perturbation drives the instability.

Our use of the variational framework presented in \citep{bergliaffa2004wave} enables us to make the connection to analogous instabilities around rotating black holes.
When the vortex throat is open and the fluid is irrotational, the negative mode arises due to the presence of the analogue ergoregion, which is the defining feature underpinning black hole superradiance.
Notably, the presence of a rotational core does not preclude this mechanism; the surface instability of the vortex can still be of the superradiant type in a fluid that is not everywhere irrotational. According to our model, it is only
once the vortex throat closes and the dry patch disappears that the dominant contribution to the instability shifts to vorticity. These findings indicate that a vortex with an open throat constitutes a promising analogue system for studying superradiant instabilities.

In a broader context, superradiant instabilities have attracted considerable attention in gravitational wave astronomy because they may provide observable signatures of new physics. In particular, an instability of the black hole bomb type has been proposed to occur around astrophysical black holes for ultra-light bosonic fields such as the hypothetical axion~\citep{Detweiler:1980uk,Dolan:2007mj,Brito:2015oca}, producing a distinctive gravitational wave signal that could reveal physics beyond the Standard Model~\citep{brito2017gravitational}. The eventual outcome of this instability, however, can be profoundly modified by nonlinear interactions~\citep{yoshino2012bosenova,kodama2012axiverse,fukuda2020aspects}. Exploring the nonlinear development of the black hole bomb mechanism in analogue systems may therefore provide valuable guidance at the interface between gravitational and fundamental physics. Runaway exponential growth associated with the black hole bomb has recently been observed in an electromagnetic analogue experiment~\citep{cromb2025}.
In classical fluids, the analogue black hole bomb is closely related to the rotating polygons instability studied by~\citep{mougel2017instabilities}, although in the present terminology this is more accurately described as a hybrid instability owing to the absence of a dissipative mechanism within the vortex core. Such dissipation can be introduced by employing a draining vortex, which possesses an analogue horizon for long-wavelength surface waves~\citep{visser1998acoustic}. Recent progress in generating these flows in superfluid $^4$He~\citep{yano2018observation,matsumura2019observation,svancara2024rotating,smaniotto2025}, where the assumption of inviscid dynamics is well satisfied, offers a promising route to studying the unimpeded growth and nonlinear development of such instabilities.


\begin{bmhead}[Funding]
S.P. acknowledges support from the Engineering and Physical Sciences Research Council through the Stephen Hawking Postdoctoral Fellowship (EP/Z536660/1). S.P., P.S., P.\v{S}., S.W. and R.G.  acknowledge the Science and Technology Facilities Council for their generous support within Quantum Simulators for Fundamental Physics (ST/T006900/1 and ST/T005858/1), as part of the UKRI Quantum Technologies for Fundamental Physics programme. L.S. and S.W. gratefully acknowledge the support of the Leverhulme Research Leadership Award (RL-2019-020). M.R. acknowledges partial support from the Conselho Nacional de Desenvolvimento Científico e Tecnológico (CNPq, Brazil, grant 315991/2023-2), and from the São Paulo Research Foundation (FAPESP, Brazil, grants
\mbox{2022/08335-0} and \mbox{2025/02701-3}). S.W. also acknowledges the Royal Society University Research Fellowship (UF120112). S.W and R.G. acknowledge support from the Perimeter Institute. Research at Perimeter Institute is supported by the Government of Canada through the Department of Innovation, Science and Economic Development Canada and by the Province of Ontario through the Ministry of Research, Innovation and Science.
\end{bmhead}

\begin{bmhead}[Data availability]
Numerical codes and datasets that support the findings of this article are openly available~\citep{data}.
\end{bmhead}

\begin{bmhead}[Declaration of interests]
The authors report no conflict of interest.
\end{bmhead}

\begin{appen}

\section{Non-draining vortex flow} \label{app:viscosity}

It is essential to establish the consistency between the variational framework for inviscid fluids introduced in \S~\ref{sec:lagrangian} and the assumption of a non-draining rotational flow. To accomplish this, we  examine a more general fluid flow, characterized by its viscosity $\nu$, which satisfies the Navier-Stokes equation. We consider standard cylindrical coordinates $(r,\theta,z)$ and assume a stationary and axisymmetric free surface flow defined by $z=h_0(r)$. The velocity field on the free surface is,
\begin{equation}
    \mathbf{v}_0 = v_r(r)\hat{\mathbf{e}}_r + v_\theta(r)\hat{\mathbf{e}}_\theta,
\end{equation}
and the corresponding vorticity is 
\begin{equation}
    \bm{\zeta}_0 = \zeta_0(r)\hat{\mathbf{e}}_z.
\end{equation}
The radial component of the velocity, the viscosity $\nu$ and the vorticity scalar $\zeta_0$ can be related through the angular component of the Navier-Stokes equation,
\begin{equation} \label{NS_theta}
    v_r\zeta_0 = \nu\partial_r\zeta_0.
\end{equation}
The equation above demonstrates that, in draining vortices, the radial flow is usually compensated by viscosity to maintain a rotational core of a particular size -- see, e.g., the Burgers vortex \citep{lautrup2011physics}. 
On the other hand, for inviscid fluids, the equation above shows that the presence of a rotational core is compatible with a non-draining flow, as we have considered throughout this work.

\section{WKB method for potential flow} \label{app:wkb}

\begin{figure}
\centering
\includegraphics[scale = 0.8]{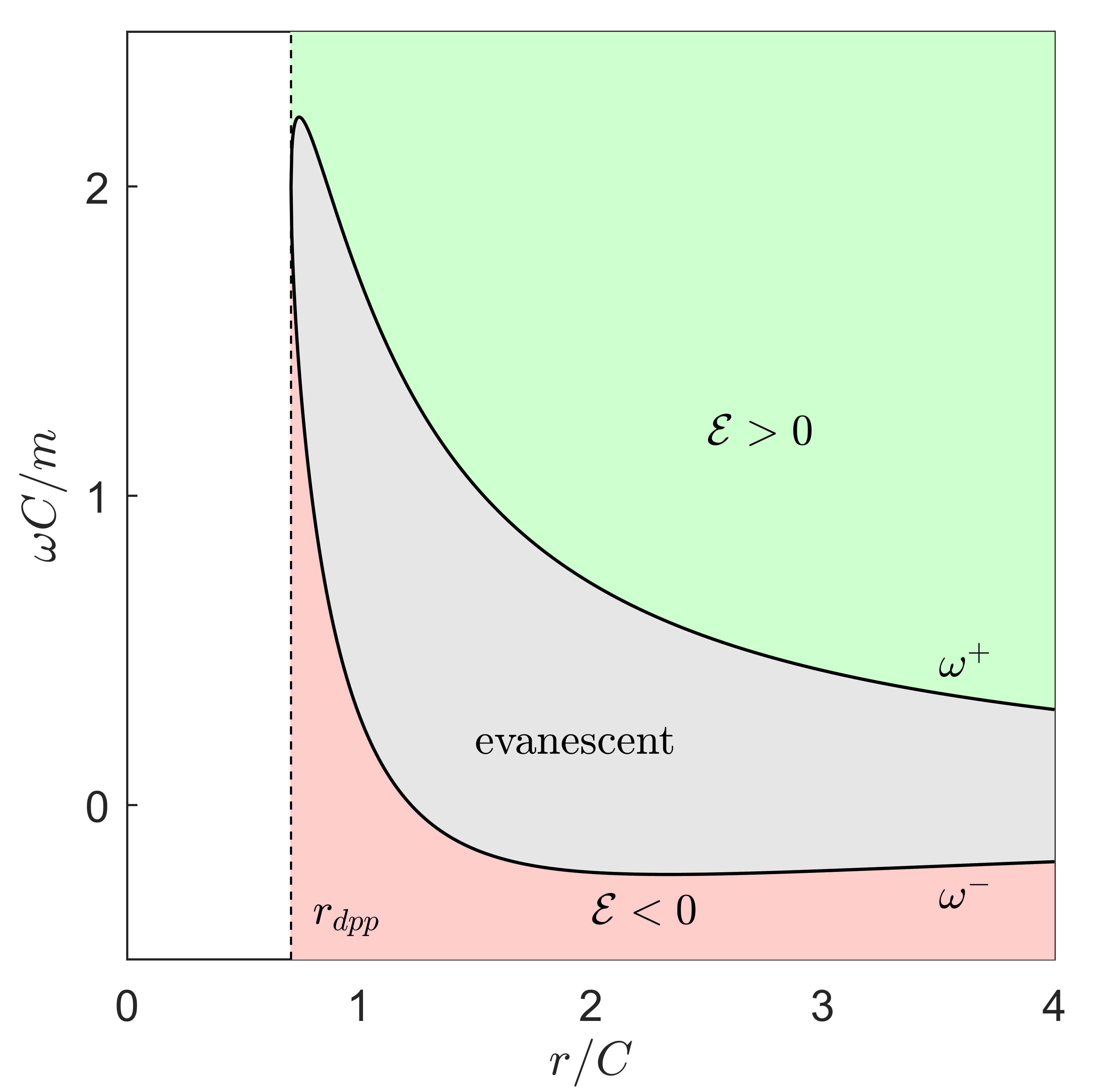}
\caption{The frequencies $\omega^{\pm}$ as a function of radius. The WKB energy density,  $\omega\mathcal{E}$, is positive in the green region and negative in the red region. Waves are evanescent in the grey area.} \label{fig:appB1}
\end{figure}

The WKB method \citep{patrick2025quasinormal} is applied to the DPP regime to show explicitly that the negative energy vortex mode is (in this case) a trapped wave inside the ergoregion.
The WKB solution of \eqref{1d_eqn} is,
\begin{equation} \label{wkbsol}
    \psi_1 = \frac{1}{\sqrt{rhp}}\left(\alpha^+e^{i\int p dr} + \alpha^-e^{-i\int p dr}\right),
\end{equation}
where $\alpha^\pm$ are the amplitudes of the ingoing and outgoing waves and the radial wavenumber $p$ is the solution of the local dispersion relation. 
In terms of the variables defined in \eqref{adim}, this wavenumber is given by
\begin{equation}
    p = \sqrt{\frac{\Omega^2}{h}-\frac{m^2}{r^2}},
\end{equation}
which has turning points ($p=0$) at $r=r^\pm_\tp$, with
\begin{equation}
    r^\pm_\tp =  \sqrt{\frac{3mC^2}{2\omega C+m \mp \sqrt{(2\omega C+m)^2-6\omega^2C^2}}}.
\end{equation}
The definition of ingoing and outgoing is determined by the sign of the radial group velocity $(\partial_\omega p)^{-1}$.

It is useful to define the following functions,
\begin{equation}    \omega^\pm(r) =  \frac{mC}{r^2} \pm \sqrt{\frac{hm^2}{r^2}},
\end{equation}
such that $\omega = \omega^\pm(r_\tp)$, i.e.
the intersection of $\omega$ with these curves gives the location of the turning points.
The significance of these curves is that if $\omega>\omega^+$ $(\omega<\omega^-)$ then $\Omega>0$ ($\Omega<0$) while if $\omega^-<\omega<\omega^+$ then $p\in\mathbb{C}$ and waves are evanescent.
Example curves are shown in Fig.~\ref{fig:appB1}.
The norm \eqref{energy} in the WKB approximation is $\mathcal{E}=\Omega|\psi_1|^2$, hence, waves above (below) $\omega^+$ ($\omega^-$) have positive (negative) norm.
A key thing to stress is that $\omega^-$ becomes positive at the outer boundary of the ergoregion $r_{e}^+$.
The energy density is $\omega \mathcal{E}$ and, hence, positivity of $\omega^-$ for certain $r$ allows the existence of propagating waves with negative energy densities. 
If there is a trapped wave in a region where $\omega^->0$, this wave is trapped inside the ergoregion and, consequently, its total energy is negative.

To compute an approximate resonance condition for trapped waves, we need to convert the regularity requirement at $r_\dpp$ into a statement about the WKB amplitudes.
We can expand the wavenumber $p$ around the boundary of the dry patch as
\begin{equation}
    p = \frac{|\Omega_\dpp|}{\sqrt{h'_\dpp(r-r_\dpp)}} + \mathcal{O}(\sqrt{r-r_\dpp}),
\end{equation}
where $\Omega_\dpp=\Omega(r_\dpp)$ and $h'_\dpp=h_0'(r_\dpp)$. Inserting the expression above into \eqref{wkbsol} yields
\begin{equation} \label{wkbsol_x0}
\begin{split}
    \psi_1 \overset{r\to r_\dpp}{\sim} & \ \frac{\alpha^+_\dpp}{(r-r_\dpp)^\frac{1}{4}} e^{2i|\Omega_\dpp|\sqrt{(r-r_\dpp)/h_\dpp'}} \\
    + & \ \frac{\alpha^-_\dpp}{(r-r_\dpp)^\frac{1}{4}}e^{-2i|\Omega_\dpp|\sqrt{(r-r_\dpp)/h_\dpp'}},
\end{split}
\end{equation}
where the $\alpha_\dpp$ are the same as those in Eq.~\eqref{wkbsol} when the lower limit of the phase integral is set to $r_\dpp$. 
Next, we compare this expression with the local solution at the inner boundary.
First we expand Eq.~\eqref{1d_eqn} in powers of $r-r_\dpp$,
\begin{equation}
    (r-r_\dpp)^2\partial_r^2\psi_1 + \partial_r\psi_1 + \frac{\Omega^2_\dpp}{h'_\dpp}\psi_1 = 0,
\end{equation}
which has (up to an overall constant) the regular solution,
\begin{equation}
     \psi \propto J_0\left(2\Omega_\dpp\sqrt{\frac{r-r_\dpp}{h'_\dpp}}\right).
\end{equation}
Comparing the asymptotic expansion of this solution with \eqref{wkbsol_x0}, we find that the WKB amplitudes satisfy $\alpha_\dpp^-=i\alpha_\dpp^+$.
This is the same type of boundary condition that arises in the core of a quantum vortex \citep{patrick2022quantum}, although in that case the inner boundary condition is applied at $r=0$ rather than $r_\dpp$.
The resonance condition that arises for a wave trapped in the low $r$ region is
\begin{equation} \label{resI}
    I \equiv \int^{r^-_\tp}_{r_\dpp} p dr = \pi\left(n+\frac{1}{2}\right) \quad \Rightarrow \quad \cos (I) = 0.
\end{equation}
Note that, for $\omega< mC/r_\dpp^2$, $r^-_\tp$ is the solution of $\omega = \omega^-$.
That is, $r_\tp^-$ lies in the ergoregion for $\omega>0$.
Thus, \eqref{resI} describes the trapping of a wave inside the ergoregion.
We can then solve \eqref{resI} over the frequency range that admits negative energy densities, i.e. $\omega\in[0,mC/r_\dpp^2]$.
We find that $\cos(I)$ has no zeroes for $m=1$, in agreement with the results in \S~\ref{sec:num_int} that there are no unstable modes for $m=1$.
For $m=2$, there is a zero at $\omega C = 0.844$, which is consistent with the value of $0.849$ found for $\mathrm{Re}[\omega C]$ in the DPP regime of Fig.~\ref{fig:4}.
Similar analyses can be performed for higher $m$.
We conclude that the ergoregion is the underlying cause of negative energies in the DPP regime.

\section{Continued fraction method} \label{app:cfm}
Leaver's method~\citep{Leaver:1985ax}, also known as the continued fraction method, is applied to compute the eigenfrequencies of the Rankine vortex~\citep[see e.g.][for implementations of the method in the context of hydrodynamical vortices]{Cardoso:2004fi,
oliveira2014ergoregion, oliveira2018ergoregion, 
patrick2018black,oliveira2024ergoregion}. Using the rescaled variables defined in \eqref{adim}, we employ the ansatz 
\begin{equation} \label{cfansatz}
\psi_1 = \frac{e^{i \omega r}}{\sqrt{r}} \sum _{n=0}^{\infty} a_n \left(\frac{r - \alpha}{r- \beta} \right)^n,
\end{equation}
with $a_n \in \mathbb{C}$. We fix $\{\alpha,\beta\}=\{C/\sqrt{{2}},0\}$ in the DPP case and $\{\alpha,\beta\}=\{1,C/\sqrt{{2}}\}$ in the DPR and WP cases. These choices guarantee that, according to Fuchs' theorem, the series \eqref{cfansatz} converges everywhere in the range $r \in [\alpha,\infty)$, where the flow is irrotational. Convergence in the asymptotic limit $r \rightarrow \infty$, however, is not guaranteed \textit{a priori}. We remark that the \textit{ansatz} \eqref{cfansatz} is compatible with the appropriate boundary conditions at $r=\alpha$, as discussed in \S~\ref{sec:boundarycs}. The \textit{ansatz} above also satisfies the boundary condition of the open problem, i.e.~Eq.~\eqref{asymp} with $A^-_\infty = 0$, if the series $\sum a_n$ converges. Leaver's method relies on the fact that this series converges if and only if $\omega \in \mathbb{C}$ satisfies a certain algebraic equation involving an infinite continued fraction. 

Let us first consider the DPP case. Substituting \eqref{cfansatz} into the wave equation \eqref{1d_eqn} yields a four-term recurrence relation for the series coefficients $a_n$:
\begin{subequations} \label{recurrence1}
\begin{align} 
\alpha_0  a_{1} + \beta_0 a_0 &= 0, \label{rec1_eq1}\\
\alpha_1  a_{2} + \beta_1 a_1 + \gamma_1 a_{0} &= 0, \label{rec1_eq2} \\
\alpha_n  a_{n+1} + \beta_n a_n + \gamma_n a_{n-1} + \delta_n a_{n-2} &= 0, \quad  n \ge 2, \label{rec1_eq3} 
\end{align}
\end{subequations}
where $\alpha_n$, $\beta_n$, $\gamma_n$, and $\delta_n$ are complex coefficients. The explicit form of these coefficients, in terms of the flow parameter $C$, the azimuthal number $m$ and the frequency $\omega$, is 
\begin{subequations}
\begin{align}
&\alpha_n = 8 (1+n)^2, \\
&\beta_n = -20 n (n+1)  +4 i \sqrt{2} C \omega (2 n+1) + 2 C^2 \omega ^2-8 C m \omega +8 m^2-4, \\
&\gamma_n = 16 n^2 -4 i \sqrt{2} C n \omega -24 m^2-6, \\
&\delta_n = -4 n(n-1) + 3 + 12 m^2.
\end{align}
\end{subequations}

Eqs.~\eqref{recurrence1} can be transformed into a three-term recurrence relation of the form 
\begin{subequations} \label{recurrence1gauss}
\begin{align} 
    \alpha'_0  a_{1} + \beta'_0 a_0 &= 0, \\
    \alpha'_n  a_{n+1} + \beta'_n a_n + \gamma'_n a_{n-1} &= 0, \quad  n \ge 1,  
\end{align}
\end{subequations}
through Gaussian eliminations. The algorithm employed to obtain the primed coefficients in terms of the unprimed ones can be found, e.g., in \citet{Leaver:1990zz,Konoplya:2011qq,Richartz:2015saa}.  
Manipulation of \eqref{recurrence1gauss}, as in \citet{Leaver:1985ax}, yields the infinite continued fraction equation
\begin{equation} \label{cf_equation}
	0=\beta'_{0}-\frac{\alpha'_{0} \gamma'_{1}}{\beta'_{1}-}\frac{\alpha'_{1} \gamma'_{2}}{\beta'_{2}-}\frac{\alpha'_{2} \gamma'_{3}}{\beta'_{3}-} \ldots,
\end{equation}
which must be solved for the complex eigenfrequencies. More precisely, given $C$ and $m$, we truncate the continued fraction at a specified order $N \ge 1$ and use a root-finding algorithm to determine $\omega$. We choose the truncation order $N$ by requiring that the relative difference between the eigenfrequencies produced with $N$ and $N+100$ terms is smaller than $10^{-5}$. Typically, $N=200$ is sufficient. We also note that the method is sensitive to the initial guess chosen for the root-finding algorithm. For a fixed $m$, we track each eigenfrequency by incrementing the value of $C$ to $C+ \delta C$, with $\delta C \ll C$ (typically $\delta C = 10^{-2}$), and using the output of the root finder for $C$ as the initial guess for $C+\delta C$.     

The application of the continued fraction method in the DPR and WP cases is similar. The main difference is that the substitution of \eqref{cfansatz} into the wave equation \eqref{1d_eqn} yields a six-term recurrence relation instead of a four-term recurrence relation,
\begin{subequations} \label{recurrence2}
\begin{align} 
\overline{\alpha}_1 a_{2} + \overline{\beta}_1 a_1 + \overline{\gamma}_1 a_{0} &= 0,\\
\overline{\alpha}_2 a_{3} + \overline{\beta}_2 a_2 + \overline{\gamma}_2 a_{1} + \overline{\delta}_2 a_{0} &= 0,\\
\overline{\alpha}_3 a_{4} + \overline{\beta}_3 a_3 + \overline{\gamma}_3 a_{2} + \overline{\delta}_3 a_{0} + \overline{\epsilon}_3 a_{0} &= 0,\\
\overline{\alpha}_n a_{n+1} + \overline{\beta}_n a_n + \overline{\gamma}_n a_{n-1} + \overline{\delta}_n a_{n-2} + \overline{\epsilon}_n a_{n-3} + \overline{\kappa}_n a_{n-4} &= 0, \quad n \ge 4.
\end{align}
\end{subequations}
The explicit form of the coefficients, for $n \ge 1$, is 
\begin{subequations}
\begin{align}
&\overline{\alpha}_n = -8 \left(\sqrt{2} C+2\right) n (n+1), \\
&\overline{\beta}_n = Q_1 n + Q_2 n^2, \\
&\overline{\gamma}_n = Q_3 + Q_4 n + Q_5 n^2, \\
&\overline{\delta}_n = Q_6 + Q_7 n + Q_8 n^2, \\
&\overline{\epsilon}_n =Q_9 + Q_{10} n + Q_{11} n^2, \\
&\overline{\kappa}_n = 8 \sqrt{2} C^3 (n-4)^2,
\end{align}
\end{subequations}
\newpage\noindent
with
\begin{subequations}
\begin{align}
&Q_1=   -16 \left(C^2 (2-i \omega ) +2 \sqrt{2} C+2 i \omega \right), \\
&Q_2=   16 \left(C^2+3 \sqrt{2} C+2\right), \\
&Q_3=   \sqrt{2} C^3 \left(12 m^2+4 \omega  (\omega +6 i)-21\right)
        -2 C^2 \left(12 m^2+8 \sqrt{2} m \omega +4 \omega^2 -8 i \omega+75\right) - \nonumber\\&\qquad
        -2 \sqrt{2} C \left(4 m^2+32 i \omega +31\right) + 32 C m \omega +16 m^2-4, \\
& Q_4=  4\sqrt{2} C^4 (5-4 i \omega )
        + 4C^2 (58 -8 i \omega )
        + 8 \sqrt{2} C (17+8 i \omega ) + 16, \\ 
&Q_5=   -4 \left( \sqrt{2} C^3 + 18 C^2 +18 \sqrt{2} C +4\right), \\
&Q_6=   -8 C^4 \omega  (\omega +3 i) + 2C^3 \left(16 m \omega +\sqrt{2} (69 -12  m^2 + 4 \omega^2  -32 i \omega)\right) +
        \nonumber\\&\qquad
        +8 C^2 \left(8 m^2-4 \sqrt{2} m \omega +22 i \omega +53\right)
        +4 \sqrt{2} C \left(17 - 4m^2\right), \\
& Q_7=  8 i C^4 \omega + 32 \sqrt{2} C^3 (i\omega - 3) - 16 C^2 ( 5i \omega + 26) -96 \sqrt{2} C , \\
&Q_8=   16 \left( \sqrt{2} C^3 +6 C^2 +2 \sqrt{2} C\right) , \\
&Q_9=   -8 C^2 \left(2 m^2+35\right)+ 2\sqrt{2} C^5 \omega ^2 
        -4 C^4 \omega  \left(2 \sqrt{2} m+\omega -14i\right) +
        \nonumber\\&\qquad
        +4C^3 \left( 2 \sqrt{2} m^2+4 m \omega  -61 \sqrt{2} -14 i \sqrt{2} \omega) \right), \\
&Q_{10}= 4 C^2 \left(-4 i C^2 \omega + \sqrt{2} C (35+4 i \omega )+54\right) , \\
&Q_{11}= -20 C^2 \left(\sqrt{2} C+2\right).
\end{align}
\end{subequations}
To obtain a continued fraction equation similar to \eqref{cf_equation}, one needs to complement Eq.~\eqref{recurrence2} with the relation
\begin{equation}
\overline{\alpha}_0  a_{1} + \overline{\beta}_0 a_0 = 0, 
\end{equation}
where 
\begin{subequations}
\begin{align}
&\overline{\alpha}_0 = \sqrt{2} C -2 , \\
&\overline{\beta}_0 = 4 \left(1-2 i \omega + 2 \frac{\psi_1'(1^+)}{\psi_1(1^+)} \right)^{-1}.
\end{align}
\end{subequations}
The ratio between $\psi_1'(1^+)$ and $\psi_1(1^+)$, obtained from \eqref{matching}, is given by 
\begin{equation}
\frac{\psi_1'(1^+)}{\psi_1(1^+)} = \frac{(\omega - m C)^2}{(\omega - m C)^2-4 C^2}  \left(\frac{\psi_1'(1^-)}{\psi_1(1^-)}-\frac{2 m C}{\omega - m C}\right),
\end{equation}
with $\psi_1'(1^-)$ and $\psi_1(1^-)$ determined by  \eqref{psi1_wp} in the WP case and by \eqref{psi1_dpr} in the DPR case.

As in the DPP regime, applying successive Gaussian eliminations to \eqref{recurrence2} yields a three-term recurrence relation of the form  
\begin{subequations} \label{recurrence2gauss}
\begin{align} 
\overline{\alpha}'_0  a_{1} + \overline{\beta}'_0 a_0 &= 0, \\
\overline{\alpha}'_n  a_{n+1} + \overline{\beta}'_n a_n + \overline{\gamma}'_n a_{n-1} &= 0, \quad  n \ge 1,  
\end{align}
\end{subequations}
where, once again, the algorithm employed to obtain the primed coefficients in terms of the unprimed ones can be found, e.g., in~\citet{Leaver:1990zz,Konoplya:2011qq,Richartz:2015saa}. The recurrence relation above then leads to an infinite continued fraction similar to \eqref{cf_equation}. 

The eigenfrequencies of the open problem for $m=2$, $m=3$ and $m=4$, calculated through the continued fraction method, are displayed in Fig.~\ref{fig:8} for rotation parameters $0 \le C \le 4$. The results, discussed in \S~\ref{sec:m34}, are compatible with the ones discussed in \S~\ref{sec:open} and Appendix~\ref{app:wkb}, obtained through the direct integration method.

\end{appen}\clearpage

\bibliographystyle{jfm}
\bibliography{biblio}

\end{document}